 \documentclass[authoryear,preprint,review,12pt]{elsarticle}
\usepackage{color}
 \usepackage{graphics}

\usepackage{amssymb}
\usepackage{rotating}
\usepackage{lineno}
\usepackage{setspace}

\begin{document}



\noindent
{\bf{\large{Integration of natural data within a numerical model of ablative subduction: A possible interpretation for the Alpine dynamics of the Austroalpine crust.}}}
\vskip 1.5 cm
\noindent
Manuel Roda

{\it{Universit\`a degli Studi di Milano, Dipartimento di Scienze della Terra ``A. Desio'', now at Universiteit Utrecht, Faculty of Geosciences,
Budapestlaan 4, 3584 CD Utrecht, THE NETHERLANDS

Corresponding author. E-mail: jediroda@gmail.com}}

\vskip 0.5 cm
\noindent
Maria Iole Spalla

{\it{Universit\`a degli Studi di Milano, Dipartimento di Scienze della Terra ``A. Desio'', Sezione di Geologia, Via Mangiagalli 34, 20133 Milano - ITALY

IDPA, CNR, Via Mangiagalli, 34, I-20133 Milan, Italy}}

\vskip 0.5 cm
\noindent
Anna Maria Marotta

{\it{Universit\`a degli Studi di Milano, Dipartimento di Scienze della Terra ``A. Desio'', Sezione di Geofisica, Via L. Cicognara 7, 20129 Milano - ITALY

IDPA, CNR, Via Mangiagalli, 34, I-20133 Milan, Italy}}

\vskip 1.5 cm
\noindent
\textbf{Short title}:
Model of ablative subduction in the Alps.	

\linenumbers

\clearpage

\noindent
{\bf{ABSTRACT}}

\noindent
A numerical modelling approach is used to validate the physical and geological reliability of the ablative subduction mechanism during Alpine convergence in order to interpret the tectonic and metamorphic evolution of an inner portion of the Alpine belt: the Austroalpine Domain. The model predictions and the natural data for the Austroalpine of the Western Alps agree very well in terms of $P$--$T$ peak conditions, relative chronology of peak and exhumation events, $P$--$T$--$t$ paths, thermal gradients and the tectonic evolution of the continental rocks. These findings suggest that a pre-collisional evolution of this domain, with the burial of the continental rocks (induced by ablative subduction of the overriding Adria plate) and their exhumation (driven by an upwelling flow generated in a hydrated mantle wedge) could be a valid mechanism that reproduces the actual tectono-metamorphic configuration of this part of the Alps. There is less agreement between the model predictions and the natural data for the Austroalpine of the Central-Eastern Alps. Based on the natural data available in the literature, a critical discussion of the other proposed mechanisms is presented, and additional geological factors that should be considered within the numerical model are suggested to improve the fitting to the numerical results; these factors include variations in the continental and/or oceanic thickness, variation of the subduction rate and/or slab dip, the initial thermal state of the passive margin, the occurrence of continental collision and an oblique convergence.

\noindent
{\bf{Key words}}: ablative subduction; numerical modelling; hydrated mantle wedge; European Alps; Austroalpine Domain.

\clearpage
\noindent
{\bf{INTRODUCTION}}

\noindent
The exhumation of $HP$--$UHP$ continental rocks is a topic that has intrigued many geoscientists since the late 1960's. Based on depth-time paths analysis and analogue and numerical modelling, pre-collisional and syn- to post-collisional exhumation mechanisms have been proposed \citep[e.g.][]{Duchene1997}. In the first case, exhumation occurs in an accretionary wedge \citep[e.g.][]{Yamato2007}, in regions of corner flow \citep[e.g.][]{Cloos1982,Shreve1986,Cloos1988a} and in the serpentinised mantle channel (wedge) \citep[e.g.][]{Gerya2005_03,Meda2010,Roda2010a}. Crust-mantle delamination \citep[e.g.][]{Chemenda1995}, continent or micro-continent collision \citep[e.g.][]{Gerya2008a,Warren2008}, slab break-off  \citep{Ernst1997}, slab retreat \citep{Ring2003}, rollback slab \citep{Brun2008}, intra-continental subduction \citep[e.g.][]{Schuster1999,Thoeni2006,Janak2009,Stuwe2010} and vertical extrusion leaded by rigid mantle indentation \citep{Rolland2000,Schwartz2007,Schreiber2010} are mechanisms that have been proposed to drive syn- to post-collisional exhumation.

Natural data and numerical predictions indicate that peculiar thermal and kinematic features characterise the uplift path of continental rocks exhumed via different mechanisms. Rocks exhumed before the collision show cold prograde and retrograde paths, metamorphic peak imprints that are mainly recorded under $HP$ and subordinate $UHP$ conditions and contrasting metamorphic evolution recorded by adjacent portions of the same structural nappe \citep{Spalla1996,Gerya2005_03,Engi2001,Meda2010}. Furthermore, pre-collisional exhumation occurs within low viscosity and/or low-density complexes, and the exhumation rate ranges from slow to moderate \citep{Roda2010a}.

Rocks exhumed during or after the onset of collision generally show warmer prograde and exhumation paths in which the retrograde trajectory can be isothermal or characterised by heating during decompression, depending on whether the uplift occurs during early or late in the collision. In these cases amphibolite facies re-equilibration is mainly predicted with the local occurrence of partial melting. The metamorphic peaks record $HP$ and $UHP$ conditions and the exhumation rates range from medium to high \citep{Ernst2008,Guillot2009}.

Although these features have been widely described in $HP$--$UHP$ orogenic belts, some controversies persist regarding the individuation of the effective exhumation mechanisms of subducted continental rocks. The debate regarding the mechanisms proposed for the exhumation of the Austroalpine continental crust in the European Alps is still open.
The Austroalpine Domain is mainly composed of continental metamorphic, igneous and sedimentary rocks derived from the distal part of the African passive continental margin (the Adria plate), involved in the Alpine subduction since the Cretaceous (early Alpine, \citet{Spalla2010} and references therein). Although there is a general recognition that these rocks have been involved in Alpine subduction-collision, their contrasting tectono-thermal evolutions and structural setting have generated different interpretations of Alpine geodynamics \citep[e.g.][]{Polino1990,Spalla1996,Beltrando2010a}, as discussed in the following section.
The identification of the exhumation mechanisms affecting the orogenic belts requires the correct evaluation of the actual dimensions of crustal and/or mantle portions that share a similar tectono-metamorphic evolution after their coupling \citep[e.g.][]{Spalla2010}. This topic remains under debate within the Alpine \citep[e.g.][]{Spalla1996,Schwartz2000,Rosenbaum2005} and extra-Alpine \citep[e.g.][]{Schneider2004,Quintero2011} scientific community.

To contribute to this debate, a numerical model of ablative subduction \citep[e.g.][]{Polino1990,Spalla1996,Meda2010,Roda2010a,Roda2011b} with hydrated mantle wedge in an ocean-continent system is performed to compare, for the first time, the results with the natural data obtained for the entire Austroalpine Domain in terms of $P$--$T$ peak estimates, relative peak chronology and exhumation time, $P$--$T$--$t$ paths, thermal gradients and the geodynamic evolution of the continental rocks. In addition, we infer the volume of those parts that likely share the same tectonic and thermal history from peak to exhumation. Other proposed mechanisms are also discussed, based on the natural data collected and a brief comparison with present-day ocean-continent subduction in North America is proposed.
\vskip 0.7 cm
\noindent
{\bf{THE AUSTROALPINE DOMAIN}}

\noindent
The uppermost tectonic domain of the Alpine belt is composed of Austroalpine continental rocks that tectonically overlie the Penninic Domain (Figs 1 and 2). The latter consists of mixed oceanic and continental slices and containing the remnants of the Tethyan domain. The western termination of the Austroalpine units is located at the northern boundary of the Lanzo Ultramafic Massif and the eastern termination coincides with the eastern border of the chain (Pannonian Basin). In the western portion of the belt, the Austroalpine rocks also occur in small slivers mingled with continental and oceanic units of the Penninic Domain.
In the following sections, the tectono-metamorphic evolution of the Austroalpine Domain will be summarised by considering three different parts (Fig. 1): (1) the Austroalpine of the Western Alps outcropping to the west of the Lepontine Dome; (2) the Austroalpine of the Central Alps between the Lepontine Dome to the west and the Tauren Window to the east; (3) the Austroalpine of the Eastern Alps that outcrops to the east of the Tauern Window.

The internal structure of the Austroalpine domain (Fig. 2), reveals a heterogeneous lithostratigraphy of the units, characterized by different metamorphic evolutions, both in Western and Eastern sectors of the chain.
The metamorphic evolution of the Austroalpine basement indicates that numerous portions, located along the entire length of the chain, have been deeply involved in subduction processes \citep[e.g.][]{Bousquet2004,Thoeni2006} because they display an eclogite- to $HP$-amphibolite-facies metamorphism, as well shown in the Map of the Metamorphic Structure of the Alps \citep{Oberhansli2004}. The $P$/$T$ ratios inferred for the eclogitic imprint generally increase from east to west, suggesting that the subduction setting of the Western Austroalpine is generally colder than the Eastern Austroalpine (Fig. 3a). However, highly-contrasting $P$/$T$ ratios can be observed in the same structural units in the entire Austroalpine Domain (i.e. the Sesia-Lanzo zone (SLZ), Fig. 3a). 
Metamorphic ages inferred for the eclogitic imprint of the Eastern Austroalpine are older than those obtained for the western part \citep[e.g.][]{Bousquet2004,Handy2004}, suggesting that the continental crust was involved in subduction in this portion of the Alpine belt earlier than in the Western Alps (Fig. 3b).
\vskip 0.5 cm
\noindent
{\bf{The Austroalpine domain of the Western Alps}}

\noindent
Alpine evolution in the Austroalpine domain of the Western Alps (AWA) is characterised by polyphase deformation under blueschist- to eclogite-facies conditions followed by retrogression under blueschist- to successive greenschist-facies conditions \citep[e.g.][]{Meda2010}. 
For the Sesia-Lanzo zone (SLZ), mineral ages ranging between 60 and 80 Ma (Fig. 3 and Table 1) have been related to the Alpine eclogite facies peak \citep[e.g.][]{Lardeaux1991,Inger1996,Duchene1997,Rubatto1998,Rubatto1999,Rebay2001,Rubatto2011}. The $P$-peak values, which are widely dispersed, range between 1.0 and 2.5 GPa, and the maximum estimated value of $T$ is approximately 650--680$^\circ$ (Fig. 4a). For the External Slices \citep{DalPiaz2001}, lower $P$-peaks have been estimated with mineral ages ranging from 40 to 50 Ma (Fig. 3 and Table 1).

Contrasting $P$--$T$ and $P$--$T$-d-t evolutions have been inferred based on phase equilibria calculations and geothermometry (\citet{Castelli2007}, and references therein) for different parts of the complex, which experienced a metamorphic imprint from eclogite to greenschists facies \citep[e.g.][]{Andreoli1976,DalPiaz1983,Kienast1983,Tropper2002,Hellwig2003,Babist2006,Rebay2007,Roda2008,Gosso2010}, Fig. 4). The $P$--$T$--$t$ paths involve high $dP/dT$ conditions and lawsonite can occur in assemblages developed during prograde or retrograde trajectories (\citet{Zucali2011}, and references therein). The estimated exhumation rates lie in the range 1.4 to 3.5 mm/a \citep{Zucali2002,Rubatto2011}. \citet{Rubatto2011} suggest that multiple stages of re-equilibration occurred under eclogite facies conditions for the rocks of the Sesia-Lanzo zone, with two subduction-exhumation cycles occurring over 20 Ma.
Large-scale shear zones developed during the final stages of greenschist facies re-equilibration in central SLZ \citep{Babist2006} and in the Dent-Blanche klippe \citep{Pennacchioni1993,Hellwig2003,Roda2008}, and a later brittle- ductile faulting postdates the Oligocene andesitic dykes and igneous stocks of the Biella and Traversella emplacement in the SLZ \citep{Zanoni2008,Zanoni2010a,Zanoni2010b}. Both plutons are emplaced in the innermost part of the SLZ (the Eclogitic Micaschist Complex) close to the internal boundary of the Alpine metamorphic units (i.e., the Periadriatic tectonic line).

Natural data related to the metamorphic peak conditions estimated for different portions of the Austroalpine of the Western Alps (reported in Figs 1a and 4a and Table 1) indicate the existence of a thermally-depressed environment compatible with active oceanic subduction (the cold subduction of \citet{Cloos1993}), with a low $T$/depth ratio also maintained during the exhumation, which is testified by the blueschist re-equilibration following the eclogite facies imprint \citep[e.g.][]{Pognante1991}. This peculiar metamorphic evolution suggests the interpretation of a burial-exhumation cycle that was accomplished during active oceanic subduction \citep[e.g.][]{Spalla1996,Zucali2004,Meda2010,Roda2011a}.

According to other authors, the occurrence of $HP$ and $UHP$ metamorphism in the Western Alps can be interpreted as episodic and induced by fast changes from shortening to extension \citep{Rosenbaum2005}. In such a tectonic scenario, orogenic processes are governed by the subduction of small ocean basins, slab rollback and the accretion of continental ribbons. In this tectonic configuration, each accretionary event is followed by revival of subduction and slab rollback in an adjacent oceanic basin, which, in turn, would switch the overriding plate into an extensional mode. The latter could then lead to the formation of metamorphic core complexes on the overriding plate and rapid exhumation of $HP$ rocks (\citet{Rosenbaum2005} and references therein).

\vskip 0.5 cm
\noindent
{\bf{The Austroalpine domain of the Central Alps}}

\noindent
Peak $P$--$T$ conditions for the eclogite facies event, in the Central Upper Austroalpine, have been derived from metabasics and their country rocks at 0.8--1.4 GPa and 500--680$^\circ$C \citep{Hoinkes1991,Poli1991,Spalla1993,Solva2001,Habler2006}. \citet{Zanchetta2007} suggests the occurrence of $UHP$ eclogites (2.6--2.9 GPa and 630--690$^\circ$C) preserved as boudins within the garnet-amphibolites (Figs 1b and 4b).
Recent geochronological data for the age of Cretaceous metamorphism in the Upper Austroalpine nappes range between 76$\pm$5 and 85$\pm$5 Ma \citep{Solva2001,Habler2006,Zanchetta2007}; an age of 143$\pm$2 Ma is inferred by \citet{Hoinkes1991} using Rb/Sr ratio on white mica (Fig. 3 and Table 2). A biotite-whole-rock Rb-Sr age of 70--80 Ma indicates that cooling below approximately 300$^\circ$C occurred during the Upper Cretaceous \citep{Habler2006}.

During the Alpine time, a clockwise $P$--$T$ path is inferred for the Err, Sella and Magna nappes: burial of the continental crust during the Ligure-Piemontese oceanic-crust subduction (between 100 and 80 Ma) and subsequent multi-stage nappe stacking under high $P$/$T$ greenschist facies conditions are interpreted to have occurred between 80 and 30 Ma \citep{Handy1996}.
\citet{Bachmann2009} indicate an age ranging from 54 to 47 Ma for the cessation of subduction induced by the Middle-Penninic (Brianzonese) collision.

With the exception of the Languard-Campo \citep{Gazzola2000} and Err/Margna nappes \citep{Handy1996,Konzett1996}, the $P$--$T$--$t$ paths for this portion record a generally higher thermal state during the early stages of the exhumation with respect to the metamorphic peak  \citep{Hoinkes1991,Habler2001a,Solva2005,Zanchetta2007}, reaching amphibolite facies conditions.

The structural and metamorphic setting of this portion of the Austroalpine Domain has led to the interpretation of a continuous tectonic erosion of the upper plate in a subduction channel context regarding the Lower Austroalpine units \citep{Bachmann2009}. Conversely, for the Upper units, a syn-collisional extruding wedge is proposed by \citet{Solva2005}.

\vskip 0.5 cm
\noindent
{\bf{The Austroalpine domain of the Eastern Alps}}

\noindent
The $P$--$T$ peak estimated for the Eastern Austroalpine exhibits a low $P$/$T$ ratio, with values of $T$ ranging from 550 to 780$^\circ$C and $P$-peak values that vary widely from approximately 1.0 to 2.5 GPa (Figs 1b and 4b).
Excluding the Sieggraben eclogites and Reckner blueschists unit, there is a marked synchronism of ages for $HP$ metamorphism along the Eastern Austroalpine of 80--90 Ma (\citet{Thoeni2006}, Fig. 3b and Table 2). The slight difference with respect to the peak-age of the Western Alps led \citet{Thoeni2006} to infer that the $HP$ event in the Eastern Alps resulted from a distinct event of continental subduction, probably related to the Meliata-Hallstatt ocean closure \citep{Kurz2003,Thoeni2006,Handy2010}. However, \citet{Handy2010} note that the Eo-alpine orogenic wedge contains no suture in the classical sense of an ophiolite belt sandwiched between metamorphosed upper and lower-plate units. The only relics of the Meliata-Hallstatt ocean consist of pelagic sediments incorporated in Jurassic-age accretionary wedges \citep{Mandl1993}.

The lack of sedimentary material sourced from high-grade metamorphic rocks in the Late Cretaceous (Campanian - Maastrichtian) central Alpine Gosau sediments may indicate a decrease of exhumation rates after approximately 85 Ma \citep{Wiesinger2005,Thoeni2006} which has been interpreted as the result of a short-lived (5--7 Ma) high-$P$ event, followed by rapid exhumation \citep{Janak2009}. This view is partly contradicted by that of \citet{Tenczer2003}, who show that northern Koralpe and the surrounding Austroalpine crystalline basement cooled to below 200$^\circ$C at 50 Ma (fission track), suggesting a lower exhumation rate than those invoked above.

The $P$--$T$--$t$ paths for this portion are characterised by a heating during the exhumation or by an isothermal uplift. Alternatively, \citet{Gaidies2008b} report an anti-clockwise Alpine $P$--$T$ path in the Rappold Complex, demonstrating a thermal state during exhumation that was colder than that of the burial event.

The exhumation of Eastern Austroalpine units could be related to the occurrence of a back-arc extension during the Ligure-Piemontese subduction successive to the Meliata closure \citep{Kurz2003} or driven by an extruding wedge and erosion during a transpressive regime within an oblique subduction event \citep{Thoeni2006,Handy2010}.

\vskip 0.7 cm
\noindent
{\bf{NUMERICAL SIMULATION}}

\vskip 0.5 cm
\noindent
{\bf{Model setup}}

\noindent
Submar code \citep{Marotta2006} is used to simulate the convergence of an ocean-continent system using a grid 1400 km wide by 708 km deep containing 6165 nodes and 3010 triangular quadratic elements as described in \citet{Roda2011a} and \citet{Roda2011b}.
The material properties and rheological laws are summarised in Table 3. Crust and mantle are compositionally differentiated via the Lagrangian particle technique \citep{Christensen1992} with a density of 1 marker (particle) per 0.25 km$^2$.
To simulate plate convergence, the velocities are fixed at the base of the oceanic crust (a depth of 10 km) with a value of 3 cm/a. To force subduction, we fixed the same velocity at depths up to 100 km at the nodes of the numerical grid distributed along a 45$^\circ$dip plane.
We assumed a continental lithosphere thickness of 80 km (with 30 km of continental crust) to represent an originally-thinned passive margin that characterised the former margin of Adria \citep[e.g.][]{DalPiaz2001a}. Using a value of approximately 40 Ma of active spreading for the Ligure-Piemontese ocean \citep{Handy2010}, we assumed 80 km thick oceanic plate  following the relation \citep{Turcotte2002}:

\begin{equation} \label{eq:age}
y=2.32(kt)^{1/2}
\end{equation}

where $y$ is the thickness of the oceanic lithosphere, $k$ is the thermal diffusivity (10$^{-6}$ $m^2/s$) and $t$ is the oceanic plate age.

The model runs for 65 Ma of active subduction at time increments of 100 ka; the model simulation ends before continental collision.
The assumed model constraints cause a strong coupling (i.e. high friction) between the upper and lower plates (see \citet{Roda2010a} for a discussion of this and additional references); therefore, our models are representative of the sole erosive margin type, which represent, at present, approximately the 60\% of the global active margins \citep{Clift2004}.

\vskip 0.5 cm
\noindent
{\bf{Model results}}

\noindent
Figure 5 illustrates the predicted temperature and velocity fields, and the distribution of markers at four time steps. In the early stages of the subduction the ablation of the overriding plate by the oceanic plate dominates the subduction dynamics. The high coupling between the two plates inhibits the formation of a large and deep trench, and only a small amount of sediment is generated (Fig. 5a). The kinematics are dominated  by the burial flow, although some small-scale upwelling flows could develop in the mantle wedge (Fig. 5a, inset). However, the intensity of the upwellings are not able to exhume a large amount of subducted material.
With the progression of mantle wedge serpentinisation (which is due to the continuous subduction and dehydration of the oceanic plate, up to 20 Ma from the beginning of subduction), a more intense upwelling flow develops, especially in the upper part of the mantle wedge (Fig. 5b). Consequently, some crustal and sedimentary material is carried up to superficial levels. The sub-horizontal flow generated along the overriding plate maintains the high degree of coupling between the two plates and induces a flexural bending of the continental plate close to the trench with consequent uprising of the lithospheric mantle. The suction generated by the high degree of coupling decreases the slab dip at depths of 200--350 km (Fig. 5b). 
During the intermediate stages of subduction the upwelling flow in the serpentinised mantle wedge becomes the dominant mechanism, the coupling between the two plates decreases and a large trench wedge develops (Fig. 5c). Large amounts of subducted materials are exhumed, and the upper and lower continental slices and sediment particles are involved in the recycling of the mantle wedge. However, only part of the recycled material reaches the surface; the rest remains in the deep portion of the mantle wedge or is more deeply buried (Fig. 5c, light purple, deep purple and yellow markers). Due to the decrease in plate coupling, suction decreases and the slab dip increases in comparison to the previous time increments. The upwelling flow increases the rising of the lithospheric mantle to subsurface levels, and thermal and mechanical erosion of the continental plate occurs.

Kinematically, the mantle wedge is characterised by three vortical flows located at depths from 30 to 100 km (Fig. 5c, inset). The late stages are characterised by a stable and highly efficient upwelling flow arising from the widespread serpentinisation of the suprasubductive mantle wedge. The consequence is a large orogenic wedge generated by the accretion of exhumed upper and lower continental crustal slices, recycled trench sediments and particles of upper oceanic crust. Even markers of continental lithospheric mantle are buried and exhumed in the mantle wedge (Fig. 5d, see markers legend). The combined effect of ablation and upwelling flow induces the thermal and mechanical erosion of the overriding plate along a zone extending from 100 to approximately 150 km from the trench, with the consequent denudation of the lower crust (Fig. 5d). The streamline pattern shows a single and intense vortical flow located in the mantle wedge from a depth of 30 to approximately 100 km (Fig. 5d, inset).

The $P$--$T$--$t$ paths followed by the upper and lower continental markers indicate burial of the material during the early stages of subduction and subsequent exhumation to subsurface levels and the upper part of the mantle wedge (Fig. 6). The mechanism of burial depends on subduction age; the $T$/$P$ ratio decreases during the intermediate stages depending on the cooling generated by the progressive subduction of the oceanic plate. Furthermore, because the trajectory of every marker could be characterised by more than one burial-exhumation loop as suggested by the $P$--$T$--$t$ paths (Fig. 6), the markers could record different values of $P_{Tmax}$ and $T_{max}$ during their evolution within the mantle wedge and different exhumation rates.

Several analyses can be performed taking the Óexhumed markersÓ into account; in this model, all particles that reach a minimum burial depth of 80 km and are subsequently exhumed to a depth of at least 30 km are considered as Óexhumed markersÓ. Analysis of the thermo-barometric conditions recorded by the exhumed markers confirms the high variability of the thermo-barometric conditions at maximum pressure ($P_{max}T$), with $P_{max}$ values from 1.2 GPa to up to 3.7 GPa, and of T, ranging from 250 to 600$^\circ$C (Fig. 7a). The distribution of the predicted thermo-barometric conditions recorded at maximum temperature by the markers ($PT_{max}$, Fig. 7b) also demonstrates high variability of the pressure associated with $T_{max}$ (from 1.0 to 3.5 GPa), but less variation in temperature, which is limited to the range 450--650$^\circ$C.
At the end of the simulation, the upper 50 km of the accretion wedge is composed of 56\% mantle and hydrated mantle, 43.5\% continental material and 0.5\% oceanic material. Markers belonging to the lower continental crust are about 17.5\% of the total amount of the exhumed material whilst 10\% represents the amount of upper continental crust.

The thermal-gradients characterising the subduction zone, which are predicted at various reference depths from the minimum values of the depressed geotherms, (Fig. 7c), exhibit the strongest decrease during the early stages of their evolution, followed by a modest decrease during the intermediate stages until a constant value is reached during the late stages. If we consider the sole mantle wedge area (up to 300 km depth), the thermal gradients during the late stages of subduction range from 2 to 4 $^\circ$C/km. Thermometric determination of $UHP$ orogenic peridotites \citep{Ernst2008} corroborate such predicted cold thermal regimes. Furthermore, the mean thermal gradients obtained from the burial path of the markers in the numerical simulation overestimate those extracted from the isotherms, at least during the first 40 Ma of subduction (Fig. 7c). Analysis of the $P/T$ ratio, which is obtained from the mean $P_{max}T$ conditions recorded by the exhumed markers at five different time steps (Figs 7d and e), confirms the progressive cooling of the mantle wedge over time during ongoing subduction; the $P/T$ values, which range from approximately 2.5 to 3.6--3.7 (10$^{-3}$ GPa/$^\circ$C), are equivalent to $T$/depth values ranging from 13 to 9 $^\circ$C/km. It is important to emphasise that these $P$/$T$ ratios do not coincide with the thermal gradients, which are exclusively represented by the configuration of the geotherms at different time steps (Fig. 7c).

We calculated the dimensions of the crustal markers groups that experienced common $P$--$T$--$t$ histories, from the peak to the exhumation (60 Ma from the beginning of the subduction), disregarding their origin (upper or lower crust). Taking into account that each marker represents a homogeneous area of 0.25 km$^2$, we found a maximum group dimension of 53 km$^2$ with a mean value of 11 km$^2$. These estimates represent the dimensions of a single metamorphic unit that means a portion of crustal and/or mantle material that shared a common tectono-thermal history within the exhumation process.

\vskip 0.7 cm
\noindent
{\bf{COMPARISON WITH NATURAL DATA}}

\noindent
In this section a comparison between model predictions and natural data from the Austroalpine continental crust is presented and the results are critically discussed. Natural data are grouped according to the locality and tectonic units, $P$--$T$ estimates and peak ages (Figs 1 and 4, Tables 1 and 2). We consider all the particles that have been buried below and recycled above a depth of 30 km (1.0 GPa), which is the starting crustal thickness, to be exhumed continental crust. For this reason, no statistical data are presented below a pressure of 1.0 GPa.

\vskip 0.5 cm
\noindent
{\bf{The Austroalpine of the Western Alps}}

\noindent
In Fig. 8, the thermo-barometric peak conditions of the natural data for the Austroalpine of the Western Alps are plotted against the numerical results in terms of of Pressure- (Fig. 8a, b) and Temperature- (Fig. 8c, d) peak conditions. Although there is in general good agreement between the $P$-peak recorded by the rocks  and that of the model, the better agreement in terms of temperature is obtained by comparison with the numerical $P$--$T$ peak conditions referred to the maximum temperature ($PT_{max}$). The predicted values of $T$ at $P_{max}$ are generally lower than those derived from the natural assemblages, particularly for $T$-values related to higher maximum pressures (from 1.5 to 2.5 GPa), whereas the $T$-values estimated for lower maximum pressures ($\leq$1.5 GPa) fit well. The comparison of natural data with model results in regard to the  $PT_{max}$ are instead ever in very good agreement. This agreement could indicate that the $P$-peak recorded by the Austroalpine rocks does not correspond to the maximum pressure reached during the rock burial, but could instead be related to the pressure conditions recorded at the maximum temperature, where the kinetics of the metamorphic reactions increase and the phase changes can be more pervasive. Conversely, this could also be due to the lack of sufficiently refined petrological methods able to identify actual pressures instead of minimal pressure conditions and reduce the uncertainty in estimates of $T$ (Fig. 4).

If the subduction simulation has a duration that is equivalent to a time-span of 65 Ma, i.e., starting at 95 Ma and ending at 30 Ma, it is possible to compare the simulation results with the timing of $P$--$T$ peak indicated in the literature. These ages range from 80 to 60 Ma for the SLZ and from 50 to 40 Ma for the External Slices (see Table 1 and Figs 2, 8e and f). 
Although older peak ages were proposed in the literature, the improvement of the dating techniques and the problem related to the excess and/or loss of Argon in the Alpine belt  \citep[e.g.][]{Ruffet1995} suggest that the more recent estimates available and not related to Ar methods are more reliable (see Table 1).
Based on geological evidence, part of the subducted material was already exhumed to shallow crustal levels by 30 Ma, and this result is compatible with the development of greenschist facies assemblages at the expense of eclogite facies ones when the subduction ceased at the beginning of continental collision \citep[e.g.][]{Platt1986,Polino1990,Zanoni2010a}. 
Following these natural contraints, the end of the model simulation (after 65 Ma, i.e., 30 Ma) is considered to broadly coincide with the age at which exhumation occurs. Based on these assumptions, there is very good agreement temporally between the simulated $P$--$T$ conditions and the record of the peak conditions in the natural rocks for both the SLZ and External Slices (Figs 8e and f). At the end of the simulation, i.e., at 30 Ma, a large number of the same continental ablated and buried markers have been exhumed to subsurface levels (Figs 8e and f, yellow markers). Some of the markers, however, reside at significant depth at the end of the simulation, and could potentially remain at depth unless exhumed by a subsequent orogenic event.
 
In Fig. 9, a comparison between selected natural and simulated $P$--$T$--$t$ paths is proposed that also takes pre-Alpine $P$--$T$ paths into account, where available in the literature. This is useful for constraining the extraction of the marker trajectories with respect to their starting structural level, at the beginning of the subduction and in the upper or lower crust, based on the results suggested by the pre-Alpine $P$--$T$--$t$ natural evolution. Furthermore, we extract the simulated $P$--$T$--$t$ paths by considering the time interval corresponding to the specific Alpine peak age proposed in the literature or, in the absence of radiometric data regarding natural peak assemblages, the $P$--$T$--$t$ path recorded during a 80 to 60 Ma time span. Although many $P$--$T$--$t$ paths are reported in the literature, we compared the numerical results with $P$--$T$--$t$ paths produced since 2000 for which modern and accurate structural and petrological analyses were available.
In Fig. 9 the grey area represents the area comprising all the numerical $P$--$T$--$t$ trajectories, and the purple and blue lines represent some examples of various paths. Orange and cyan boxes and black arrows refer to the natural $P$--$T$ conditions (pre-Alpine and Alpine) and natural $P$--$T$--$t$ paths, respectively, as given in the literature and noted in the figure legend. Natural and simulated $P$--$T$--$t$ paths are generally in good agreement in terms of both prograde and retrograde trajectories with the exception of the External Slice paths ÓArollaÓ and ÓMt. MorionÓ (Fig. 9). In these cases, as discussed in the literature, peak conditions are attained close to the onset of continental collision at approximately 30 Ma, \citep[e.g.][]{Zanoni2010a,Handy2010} and consequently the thermal regime, characterising the exhumation of these rocks, can increase due to high shear heating and strong radioactive decay  \citep[e.g.][]{Gerya2008a,Warren2008}. It must also be emphasized that our numerical simulation is stopped before the collision event, and therefore, exhumation is only partially accomplished under a sensible low-temperature regime.

\vskip 0.5 cm
\noindent
{\bf{The Austroalpine of the Central Alps}}

\noindent
The comparison between the natural and the simulated thermo-barometric conditions of the Central Alps shows no agreement with regard to $P_{max}T$ (Figs 10a and b) values because of the higher temperatures retrieved from the natural data. However, the natural data are consistent with the higher simulated $T_{max}$ values in the $PT_{max}$ diagram (Figs 10c and d). The samples that record lower peak-pressure (lower than 1.0 GPa) are not compared to the simulations.

For the comparison between the natural age data and the timing of the $P$--$T$ peak from the model, we extract the simulated $P$--$T$ values for the first 20 Ma of subduction (i.e., from 100 to 80 Ma, as suggested in the literature, Fig. 3b and Table 2), and the exhumation age corresponds to 50 to 35 Ma \citep{Handy1996,Bachmann2009}. Fig. 10e allows a qualitative estimate of fitting percentage between simulated $P$--$T$ peak values (red dots) and natural estimates (shaded area) of at least 50\%. Furthermore, the predicted $P$--$T$ conditions at the end of simulation (orange dots) are coherent with natural $P$--$T$ data set and associate age.

Only one natural $P$--$T$--$t$ path agrees well with the predictions of the model (Fig. 9: ``Mortirolo'' (ref. 1 Central Alps, Fig. 1b)). Other $P$--$T$ paths  from the literature display a warmer thermal state for both the peak and exhumation conditions and are therefore not consistent with the simulated $P$--$T$--$t$ paths.

\vskip 0.5 cm
\noindent
{\bf{The Austroalpine of the Eastern Alps}}

\noindent
A comparison of the natural and the simulated thermo-barometric conditions of the Eastern Alps shows no agreement with regard to the $P_{max}T$ values and a poor agreement with $PT_{max}$ values due to the higher thermal state suggested by the natural data (Fig. 11).
For the Eastern Alps, we considered 130 Ma to be the age at which subduction began, based on the peak age of Sieggraben (Table 2). 
In this region, in contrast to the the Western Alps, the $P$-peak ages lie in the range from 100 Ma to 90 Ma, and two exhumation stages are considered: 80 Ma \citep{Thoeni2008} and 50 Ma \citep{Tenczer2003}.

A low correspondence is obtained between the timing of the $P$-peak for the Eastern Austroalpine (90--100 Ma) and the model predictions (Fig. 11e). Regarding the peak age interval, good agreement is obtained only for lowest $P$ values. For the higher pressures, a colder thermal state is predicted by the simulation (Fig. 11e). A significant better agreement is found if the subduction beginning is assumed at 100 Ma (excluding Sieggraben), and the predicted $P$--$T$ conditions refer to the early stages of subduction (1--10 Ma = 100--90 Ma, Fig. 11f). Indeed, for this time span, a higher thermal state is predicted than for the intermediate stages of the subduction. In any case, a considerable amount of buried material has already been exhumed at both exhumation time-steps considered, although a large part sinks to greater depths and probably will remain buried.

As with the Central Austroalpine, only one natural $P$--$T$--$t$ path agrees well with the predictions of the model (Fig. 9: ``Sopron'' (ref. 19.2 Eastern Alps, Fig. 1c)). In this case, only the peak-pressure is well reproduced, whereas the thermal state of the exhumation path is underestimated by the simulated paths. Other natural $P$--$T$ paths display a warmer thermal state for both the peak and exhumation conditions and are not reproduced by the simulations.

\vskip 0.7 cm
\noindent
{\bf{DISCUSSION}}

\noindent
The model predictions agree well with the natural $P$--$T$ estimates and the $P$--$T$--$t$ paths derived from for the Austroalpine crust of the Western Alps. All natural $P$--$T$ data are reproduced by the numerical $P$--$T$ predictions, especially the $P$--$T$ peak conditions obtained by $PT_{max}$ analysis. Even the time interval between the peak and exhumation conditions are closely consistent with the  ages inferred from the rocks, with different $P$-peak ages, for the SLZ and the External Slices. In this context, the exhumation rate suggested by the numerical model ranges from 0.1 to 1 cm/a, with mean values of 0.3--0.4 cm/a. These exhumation rates agree with those predicted by \citet{Roda2010a} and suggested by \citet{Zucali2002} and \citet{Rubatto2011} for a part of the SLZ. Good agreement is found between the natural and simulated $P$--$T$--$t$ paths, both with and without consideration of their pre-Alpine evolution, with few exceptions. These cases are characterised by a peak-age very close to the collisional event, and their retrograde paths could be affected by the increase of temperature induced by the continental collision.

The same heterogeneity that characterises both the natural and predicted $P$--$T$ peak conditions and the $P$--$T$--$t$ evolutions in the same structural domain suggests the juxtaposition of rocks with different metamorphic evolutions. The rocks have been coupled and decoupled at different stages during their trajectories in the mantle wedge \citep[e.g.][]{Spalla1996,Gerya2005_03,Bousquet2008b,Roda2010a,Spalla2010}.
In this context it is evident that the tectono-metamorphic unit definition requires an evaluation of the actual size of crust and mantle-derived slices that coherently shared, after amalgamation, the same tectonic and thermal history. In light of this, the structural and metamorphic evolutions of basement rocks trace their transit throughout different levels of the lithosphere and sub-lithospheric mantle \citep[e.g.][]{Schwartz2000,Schwartz2007,Spalla2010,Quintero2011}, rather than the purely lithologic associations or the common dominant metamorphic imprint. The tectono-metamorphic unit dimensions obtained in this work are at least one order of magnitude lower than those proposed for the Alps \citep[e.g.][]{Oberhansli2004}: this means that the refinement of a method for the identification of contours of thermally-characterized and structurally-distinct units became crucial \citep{Spalla2005,Spalla2010} to obtain a good definition of the Alpine tectono-metamorphic units. In the case of the Western Austroalpine, the model results and their fitting with the natural data indicate that a nappe (Sesia-Lanzo Zone) can result from the accretion of different tectonic units.

The thermal gradient of the simulated subduction zone lies in a range from 2 to 4 $^\circ$C/km (Fig. 7c). \citet{Groppo2009}, based on the $P$--$T$ assemblages obtained for some samples belonging to the ``Lago di Cignana'' unit, suggest a gradual decrease in the thermal gradient (from 9--10 to 5--6 $^\circ$C/km) over time. This variation, which is simply related to the time, has been interpreted by these authors as evidence for an increase in the subduction rate of the Ligure-Piemontese oceanic slab during the Eocene. However, a deeper analysis of $P$--$T$--$t$ paths reveals that the attribution of thermal gradient significance to a single and transient $P$--$T$ estimate cannot be directly related to the variation of the geotherms. The inference of variation in thermal gradients requires a minimum of two $P$--$T$ estimates and the knowledge of the  configuration of isotherms in an evolving dynamic system such as an active subduction zone. Thus, the prograde path inferred by \citet{Groppo2009} suggests a thermal gradient of approximately 3 $^\circ$C/km, which agrees well with that predicted by the numerical model, rather than the transition between two different thermal gradients proposed by the authors. 
Furthermore, the numerical model suggests that the thermal gradients inferred by the burial path of the markers are quite different than those extracted from the isotherms (Fig. 7c). This finding indicates that the determination of thermal gradients that are derived directly from the $P$--$T$ path could be affected by uncertainties arising from the nonlinear trajectories of the markers in the mantle wedge and to variability in the thermal regime.

The model results suggest the occurrence of a very dynamic mantle wedge that produces two or three subduction-exhumation loops recorded by the markers within 65 Ma of active subduction (Fig. 6). This aspect of the modelling is consistent with the recent discovery of two subduction-exhumation cycles recorded in the same rocks in the Western Austroalpine of the Alps (the Sesia-Lanzo Zone \citep{Rubatto2011}. Similar subduction-exhumation cycles have been invoked to justify the complex $P$--$T$ trajectory inferred for the rocks of the Caribbean subduction channel \citep{Quintero2011} and the Franciscan complexes \citep{Wakabayashi2007,Wakabayashi2011}. 

All of the previous points suggest a pre-collisional evolution of the Austroalpine of the Western Alps, involving the burial of slices ablated by the margin of the overriding continental plate (the Adria plate) during subduction and their successive exhumation (driven by the upwelling flow generated in a hydrated mantle wedge).

A much more modest agreement between the natural data and model predictions occurs for the evolution of Austroalpine rocks from the Central and Eastern Alps. In these regions, the simulated thermal state is generally underestimated with respect to that derived from $P$--$T$ estimates of the rocks. Although the timing of simulated $P$--$T$ peak and exhumation are comparable with those suggested in the literature, only a small part of the predicted $P$--$T$ conditions is consistent with the natural conditions.

These results suggest that entirely pre-collisional ablative mechanisms are unable to explain the high temperature recorded by the rocks belonging to the Austroalpine of the Eastern Alps. A model that includes continental collision could better explain the high thermal state reached by the rocks before and during exhumation \citep{Faccenda2008,Gerya2008a,Warren2008}, although the high levels of mixing of units predicted by the cited models do not appear to reproduce the low level of heterogeneity observed in the Eastern Alps, according to \citet{Janak2009}. An intra-continental subduction zone is the suggested alternative interpretation \citep{Janak2009}. However, a mechanism able to engage an intra-continental subduction is difficult to identify. The thickening of the continental crust of the upper plate is probably a more common mechanism (which is effective during active subduction) that could be invoked to explain higher values of $T_{max}$. Indeed, the influence of a thick upper plate on the thermal state of the wedge has been described in the case of low slab dip and strong plate coupling, where a higher thermal regime characterises the mantle wedge during active subduction before continental collision \citep{Roda2011b}. This finding suggests that the higher $T$ values recorded in the Eastern Austroalpine could be related to lateral variations of upper plate thickness along the Alpine convergent margin.
As noted previously, the thermal state depends on the subduction age (early vs. intermediate subduction, Figs 7d and 11).

The poor lithologic heterogeneity assessed in the Eastern Alps is weakened by the heterogeneities in the $P$--$T$ peak conditions and $P$--$T$--$t$ evolutions of natural rocks both within single and between different units (Fig. 1 and Table 2).
The lithostratigraphic homogeneity that characterises the Eastern Austroalpine with respect to the melange-like assemblage of the Western Austroalpine does not necessarily imply homogeneous and coherent $P$--$T$--$t$--$d$ paths. The lithostratigraphy of the Western Austroalpine is comparable to that described in the Central-Eastern Austroalpine, but a high level of heterogeneity is predicted by the different $P$--$T$--$t$--$d$ evolution recorded by the slices constituting the same structural domain (or ``Alpine nappe'') \citep[e.g.][]{Schwartz2000,Bousquet2008b,Spalla2010}. In this context, better-refined prograde $P$--$T$--$t$ paths and, in general, a wide spread in derived $P$--$T$--$t$ or $P$--$T$--$t$--$d$ paths are needed for the Austroalpine of the Central-Eastern Alps to allow the existence of possible tectono-metamorphic homogeneities to be discerned.

\citet{Roda2010a,Roda2011b} suggest that a thinned oceanic plate (less than 80 km, as reasonable for a young ocean) and/or a variation of the subduction rate and/or the slab dip could affected the thermal state of the mantle wedge. Taking into account an oblique Alpine subduction and the possible V-shape of the Ligure-Piemontese ocean \citep[e.g.][]{Schmid2004}, it should be possible to obtain a decrease of the thermal state of the subduction zone from east to west that is related to an increase in the oceanic plate thickness. 

However, some natural $P$--$T$ conditions reflect a very high thermal state, partially consistent with the numerical predictions obtained for the early stages of the subduction. This observation suggests that the initial LP thermal state of the rocks could influence the thermal evolution of buried-exhumed rocks during the subduction. \citet{Marotta2009} (and references therein) indicate that the Austroalpine domain is widely affected by $HT$--$LP$ Permian-Triassic metamorphism, which developed during the early stages of the Tethyan rifting. The subduction of the Eastern Austroalpine crust earlier than that of the Western Austroalpine crust could imply a thermally perturbed continental margin, a younger (and therefore hotter) oceanic plate and a higher $T$/$P$ ratio in the early stages of the subduction. 

Moreover, horizontal movement linked to transpressive subduction could affect the $P$--$T$ conditions recorded during the exhumation paths, as noted by \citet{Thompson1997a,Thompson1997b}. In this case, the exhumation is characterized by a transition from blueschist to granulite facies conditions consequent to the increase of the transpressive component.

For the comparison between natural data from the Austroalpine domain of the Alps and the model predictions, we briefly compare the simulation results with the geodynamic scenarios proposed for still active subduction settings such as Cascades and Franciscan. In these contexts, the continental collision does not affect the subduction dynamics as in the Alps.
The thermal structure of the Cascades is not so different with respect to the simulated scenario: indeed, if we refer to fig.2 of \citet{Hyndman2005}, notable similarities between the two models can be found. In the Cascades a cold fore-arc and thermal erosion of the continental lithosphere beneath the back-arc region are present. The same thermal structure is also predicted by our model, with a cold fore-arc region almost 100 km wide and the rising of the 1400 K isotherm up to 50 km depth beneath the back-arc region, accounting for a high thermal regime. As a consequence, we do not expect differences concerning the exhumation mechanisms but a warmer exhumation-related $P$--$T$ evolutions would be consequent to the shallower position of the 1600K isotherm (asthenosphere-lithosphere boundary).
Another major difference between the two models is the larger extension of the back-arc region in the Cascades. This could be related to the longer subduction activity of the Cascades (180-160 Ma) with respect to 65 Ma in our model \citep{Currie2008}, that implies a continuous thermal erosion of the continental upper plate. Furthermore, the higher subduction dip of the Cascades \citep[medium and deep dips:][]{Lallemand2005,Cruciani2005,Currie2008} with respect to the presented model, would induce a larger back-arc hot region due to a very efficient convective cell beneath the upper plate \citep{Currie2008,Roda2010a,Roda2011b}.
The ridge very close to the trench in the Cascades scenario \citep{Hyndeman1995,Flueh1998} also implies a different thermal state with respect to that obtained by our model. Finally, the oblique subduction that characterizes large part of the Cascades could generate a perturbed thermal regime within the subduction zone \citep{Thompson1997a,Thompson1997b,Rondenay2008}.

Although an early intra-oceanic subduction scenario is invoked for the Franciscan compared with the ocean-continent setting adopted in our simulations, a lot of similarities between natural and simulated data are recognizable: (i) a wide range of peak metamorphic ages recorded by the exhumed rocks \citep{Wakabayashi2007}; (ii) different $P$--$T$ evolutions resulting from different tectonic trajectories during subduction and exhumation, with the consequent accretion of different tectono-metamorphic units at different times in the same subduction zone \citep{Wakabayashi2007}; (iii) the record of multiple (2 or 3) burial-exhumation cycles of blue-schist rocks within 65 Ma of active subduction \citep{Wakabayashi2011}, although these rocks have been related in the past to an accretionary wedge / subduction channel mechanism rather than to a hydrated mantle wedge dynamics \citep{Cloos1988a}; (iv) the presence of $HP$--$HT$ metamorphic rocks, which show anti-clockwise $P$--$T$--$t$ paths likely related to the earliest steps of the subduction zone \citep{Smart2009} that are coherent with the predicted $P$--$T$--$t$ of Fig. 6; (v) prograde eclogitization of a relatively cold subducting slab, and subsequent exhumation and blueschist-facies recrystallization of $HP$ rocks characterized by exhumation rates of several mm/a, as suggested by the numerical results \citep{Roda2010a,Roda2011a}.

\vskip 0.7 cm
\noindent
{\bf{CONCLUSIONS}}

\noindent
The general very good agreement between the predictions of the model and the natural data obtained for the Austroalpine of the Western Alps suggests that the ablation of the upper plate lithosphere coupled with the dynamics of the hydrated mantle wedge is a valid pre-collisional mechanism that can reproduce the natural burial and exhumation evolution of this portion of the Austroalpine domain during active subduction. Although this model appears to be less compatible with the natural data of the Central-Eastern Alps, a variation of the oceanic and/or upper plate thickness along the Ligure-Piemontese ocean, a variation of the subduction rate and/or the slab dip, the initial thermal state of the passive margin, the occurrence of continental collision or an oblique subduction could justify a variation in the thermal state from east to west along the Alpine convergent margin. 
Furthermore, the simulated ocean-continent dynamics allows the formation of a metamorphic complex composed of juxtaposed of rocks with different metamorphic evolutions, likely with a homogeneous lithostratigraphic setting, accreted in the mantle wedge and exhumed during active subduction.
To evaluate the compatibility between natural data and the thermo-mechanical configuration generated by alternative models, such as micro-continent collision \citep{Rosenbaum2005} and intra-continental subduction \citep{Thoeni2006,Janak2009}, numerical simulations of these models could be performed in the future using the same approach.

\vskip 0.7 cm
\noindent
{\bf{ACKNOWLEDGEMENTS}}

\noindent
We would like to thank the anonymous reviewers for their constructive criticism of the text and in particular Richard White for his accurate editorial work that greatly improved the final version. Prin 2008 ``Tectonic trajectories of subducted lithosphere in the Alpine collisional orogen from structure, metamorphism and lithostratigraphy'', CNR-IDPA and PUR 2008 ``La ricerca geofisica: esplorazione, monitoraggio, elaborazione e modellazione'' are gratefully acknowledged. Many thanks to American Journal Experts for the English revision.



\clearpage

\noindent
{\bf{FIGURE CAPTIONS}}

\noindent
Fig. 1: Locations from which natural data were obtained within the Austroalpine Domain upon the structural map of the Alps, redrawn after \citet{Schmid2004}, \citep{Thoeni2006} and \citet{Castelli2007}. Western Alps (a). Refs. 1: \citet{Castelli1991}; 2: 	\citet{Vuichard1988}; 3: 	\citet{Lardeaux1991};
4: 	\citet{Castelli2002};
5: 	\citet{Lardeaux1982};
6: 	\citet{Gosso1982};
7: 	\citet{Hy1984};
8: 	\citet{Oberhansli1985};
9: 	\citet{Compagnoni1977};
10: 	\citet{Zucali2002};
11: 	\citet{Delleani2010};
12: 	\citet{Rubatto1999};
13: 	\citet{Giorgetti2000};
14: 	\citet{Tropper2002};
15: 	\citet{Tropper1999};
16: 	\citet{Williams1983};
17: 	\citet{Spalla1997} and \citet{Gosso2010};
18: 	\citet{Ruffet1995};
19: 	\citet{Zucali2004} and \citet{Zucali2011};
20: 	\citet{Andreoli1976};
21: 	\citet{Reinsch1979};
22: 	\citet{Desmon1977};
23: 	\citet{Pognante1989a};
24: 	\citet{Pognante1987};
25: 	\citet{Rebay2007} and \citet{Lardeaux1991};
26: 	\citet{Pognante1989a};
27: 	\citet{Williams1983};
28: 	\citet{Pognante1987};
29: 	\citet{Lardeaux1983};
30: 	\citet{Pognante1987};
31: 	\citet{Pognante1989a};
32: 	\citet{Rolfo2004};
33: 	\citet{Hellwig2003};
34: 	\citet{Roda2008};
35: 	\citet{Kienast1971};
36: 	\citet{Hopfer1995};
37: 	\citet{Kienast1983};
38: 	\citet{DalPiaz1983};
39: 	\citet{Pennacchioni1988};
40: 	\citet{Scambelluri1998};
41: 	\citet{DalPiaz2001};
42: 	\citet{Cortiana1998};
43 and 44: 	\citet{Rubatto2011}. Central Alps (b). Refs. 1:	\citet{Gazzola2000};
2:	\citet{Spalla1995};
3:	\citet{Tomaschek1998};
4:	\citet{Handy1996};
5:	\citet{Guntli1989};
6:	\citet{Konzett1996};
7:	\citet{Habler2001a};
8:	\citet{Solva2001};
9:	\citet{Spalla1993};
10:	\citet{Hoinkes1991};
11:	\citet{Zanchetta2007};
12:	\citet{Habler2006}. Eastern Alps (c). Refs. 1:	\citet{Koller2003};
2:	\citet{Linner1999};
3 and 4	\citet{Faryad2003};
5:	\citet{Gaidies2008b};
6:	\citet{Gaidies2008a};
7:	\citet{Thoeni2008};
8:	\citet{Thoeni1996};
9:	\citet{Stuwe1995};
10, 11, 12 and 13	\citet{Tenczer2003};
14:	\citet{Gregurek1997};
15:	\citet{Tenczer2006};
16:	\citet{Janak2009};
17:	\citet{Putis2002};
18:	\citet{Neubauer1998};
19:	\citet{Torok1999} and \citet{Torok2003};
20:	\citet{Tropper2001}.

\vskip 0.5 cm
\noindent
Fig. 2: Cross-sections for Eastern (AA'), Central (BB') and Western (CC') Alps redrawn after \citet{Polino1990}, \citet{Dalpiaz2003} and \citet{Schuster2004}. In the Central Alps cross section the lower austroalpine comprises slices of basement units of the Err and Bernina systems, interlayered with thin slivers of Mesozoic sediments \citep[e.g.][]{Froitzheim1994}, giving rise to a complex setting characterised by the repetition of cover and basement lithostratigraphic units.

\vskip 0.5 cm
\noindent
Fig. 3: $P$/$T$ ratios and peak ages obtained from the collection of natural data (Tables 1 and 2).

\vskip 0.5 cm
\noindent
Fig. 4: Peak $P$--$T$ estimates obtained from the natural data collected for the Western (a), Central (b) and Eastern (c) Austroalpine; the numbers refer to locations marked in Fig. 1. The metamorphic facies fields shown are after \citet{Ernst2008}. GS: greenschist facies; BS: blueschist facies; EP--A: epidote-amphibolite facies; A: amphibolite facies; A--E: amphibole-eclogite facies; EP--E: zoisite-eclogite facies; Lws--E: lawsonite-eclogite facies; GR: granulite facies; HGR: kyanite-granulite facies.

\vskip 0.5 cm
\noindent
Fig. 5: Four evolutionary stages of the model: (a) 10 Ma, (b) 20 Ma, (c) 35 Ma and (d) 60 Ma. In the insets, the streamlines (blue lines) obtained in the corner area are reported. 1) Upper continental crust, 2) lower continental crust, 3) upper oceanic crust, 4) lower oceanic crust, 5) sediments, 6) exhumed upper continental crust, 7) exhumed lower continental crust, and 8) continental lithospheric mantle.

\vskip 0.5 cm
\noindent
Fig. 6: An example of a $P$--$T$--$t$ path for a lower continental marker at four time increments; the red circle indicates the initial position of the marker, and the black circle indicates the marker location at a particular subduction age. The metamorphic facies fields are from \citet{Ernst2008}. GS: greenschist facies; BS: blueschist facies; EP--A: epidote-amphibolite facies; A: amphibolite facies; A--E: amphibole-eclogite facies; EP--E: zoisite-eclogite facies; Lws--E: lawsonite-eclogite facies; GR: granulite facies; HGR: kyanite-granulite facies.

\vskip 0.5 cm
\noindent
Fig. 7: Predicted peak $P$ and $T$ conditions recorded by the exhumed markers (see the legend for the lithologic affinities) in terms of $P_{max}T$ (a) and $PT_{max}$ (b). (c): thermal gradients in the subduction zone for seven reference depths obtained from a comparison of the dynamics evolution of the isotherm configuration to the thermal gradients obtained by the burial paths of two markers (blue = upper continental crust; green = lower continental crust). (d): The $P$/$T$ ratio obtained based on $P/T$ ratio obtained by the mean $P_{max}T$ conditions recorded by the exhumed markers for five different time steps. a): Upper continental crust and b): lower continental crust.

\vskip 0.5 cm
\noindent
Fig. 8: Comparison of the natural $P$--$T$ peak assemblages coming from the Austroalpine of the Western Alps (blue dots) and simulated $P_{max}T$ and $PT_{max}$ data (grey dots) for slices ablated from upper (a-c) and lower (b-d) continental crust. e) $P$--$T$ peak assemblages obtained for the peak age of SLZ (blue dots) and f) for the External Slices (blue dots) compared to the natural estimates (shaded area). The orange dots represent the simulated $P$--$T$ data at the end of the subduction period (30 Ma), which is interpreted as the exhumation age.

\vskip 0.5 cm
\noindent
Fig. 9: Comparison of the simulated and natural $P$--$T$--$t$ paths inferred for the Austroalpine. See caption of Fig. 1 for the reference numbers. Pre-Alpine $P$--$T$--$t$ path references are 10: \citet{Lardeaux1991}; 25: \citet{Lardeaux1991} and \citet{Rebay2001}; 34: \citet{Kienast1971}, \citet{Gardien1994} and \citet{Roda2008}; 01: \citet{Gazzola2000} and 19.2: \citet{Schuster2001}.

\vskip 0.5 cm
\noindent
Fig. 10: Comparison of the natural $P$--$T$ peak assemblages coming from the Austroalpine of the Central Alps (red dots) and simulated $P_{max}$T and $PT_{max}$ data (grey dots) for slices ablated from upper (a-c) and lower (b-d) continental crust. e): $P$--$T$ peak assemblages obtained for the peak age of the Central Austroalpine (red dots) compared with the natural assemblages (shaded area). The orange dots represent the $P$--$T$ data for the period from 50 to 35 Ma, which is interpreted as the exhumation age.

\vskip 0.5 cm
\noindent
Fig. 11: Comparison of the natural $P$--$T$ peak assemblages coming from the Austroalpine of the Eastern Alps (green dots) and simulated $P_{max}$T and $PT_{max}$ data (grey dots) for slices ablated from upper (a-c) and lower (b-d) continental crust. e) and f): $P$--$T$ peak conditions predicted after 30--40 Ma from the beginning of subduction obtained for the two different starting ages of the simulation: 130 Ma (e) and 100 Ma (f) compared to natural estimates (shaded area). The red and orange dots represent the $P$--$T$ data at two times from the start of the subduction and are interpreted as two natural exhumation ages (80 Ma red; 50 Ma orange).

\clearpage
\noindent
\bf{FIGURES}
\begin{figure*}[!ht]
\includegraphics[width=14cm]{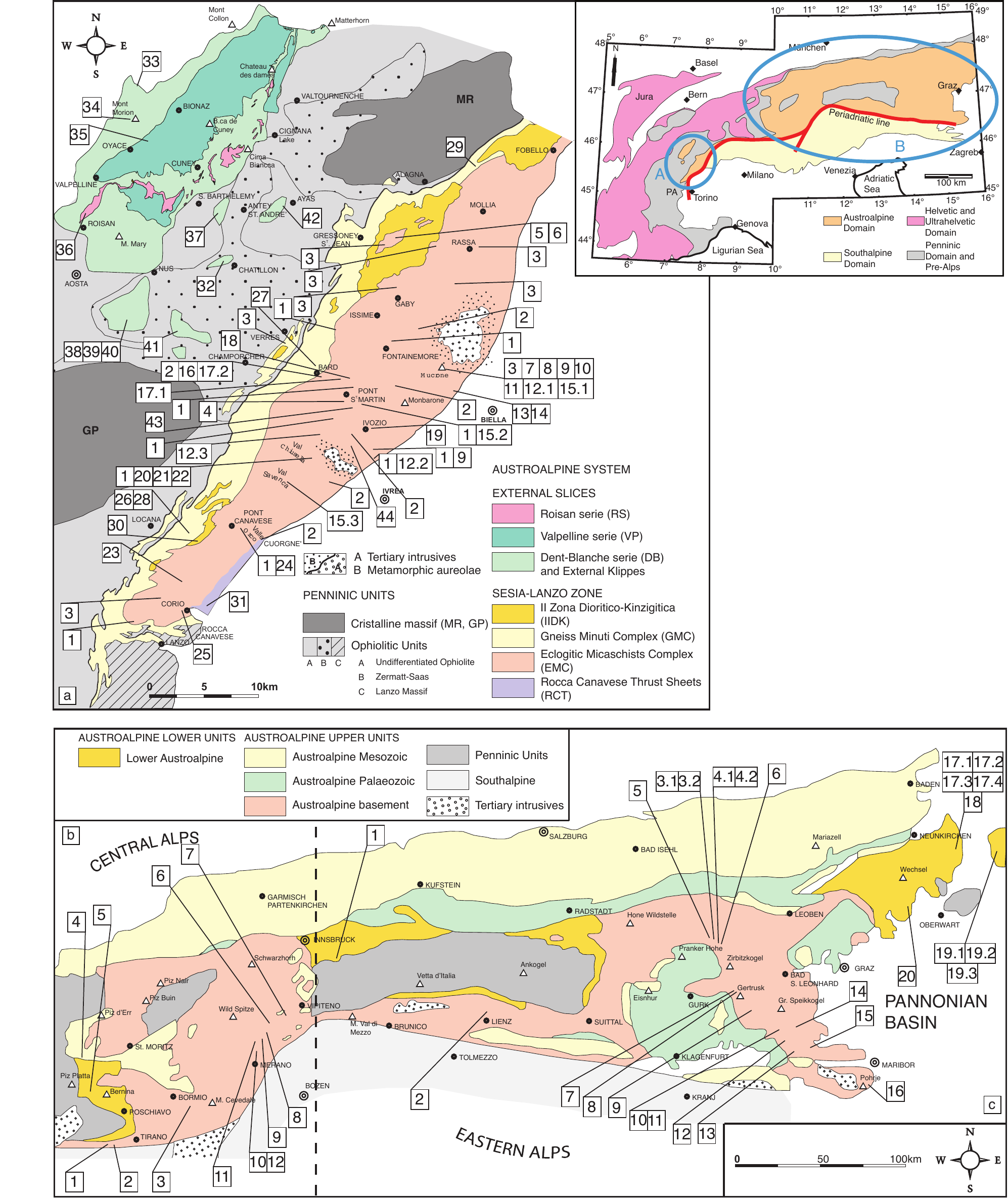}
\noindent
Fig. 1
\end{figure*}

\vskip 0.5 cm
\noindent
\begin{figure*}[!ht]
\includegraphics[width=14cm]{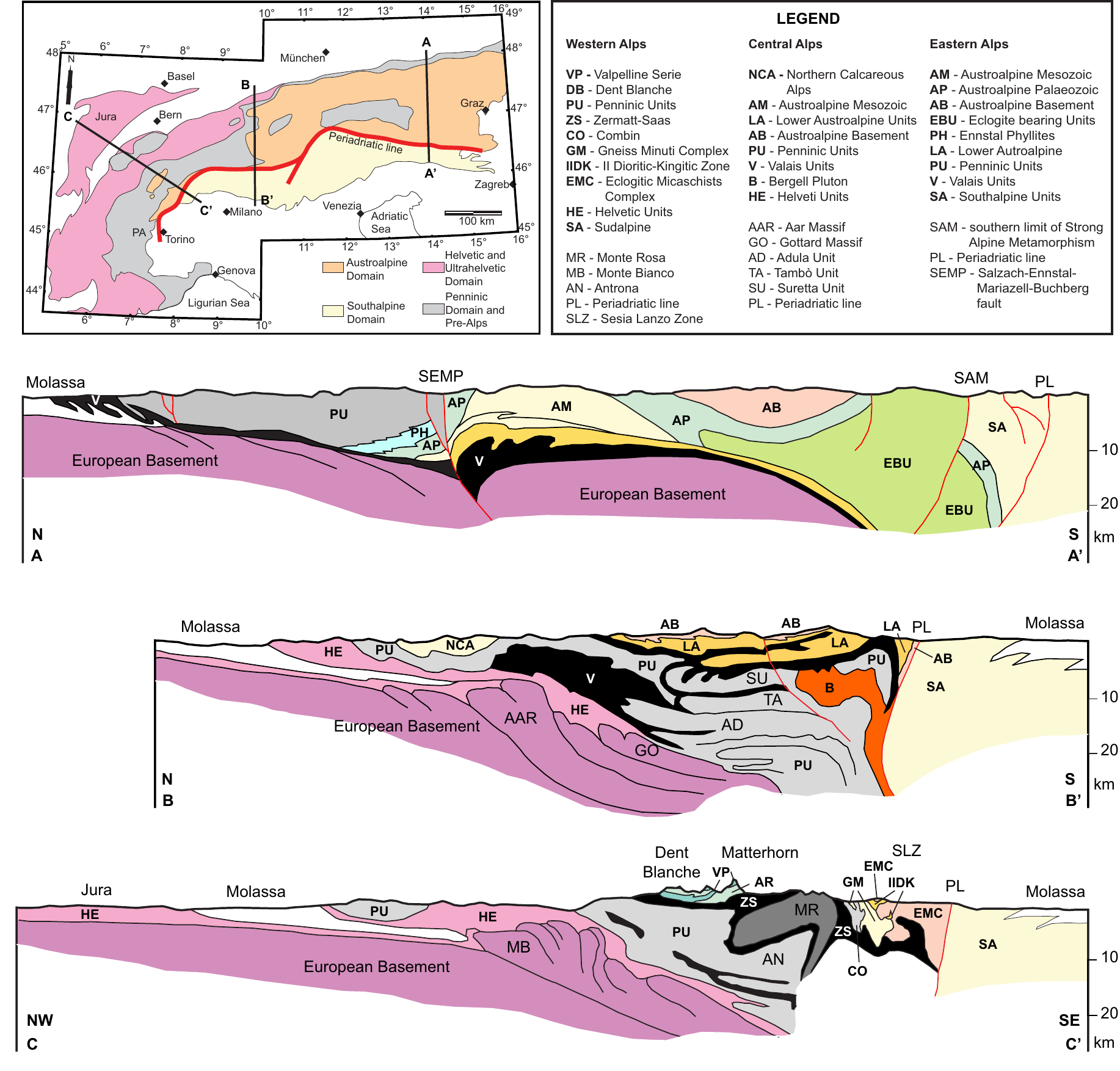}
Fig. 2
\end{figure*}

\vskip 0.5 cm
\noindent
\begin{figure*}[!ht]
\includegraphics[width=14cm]{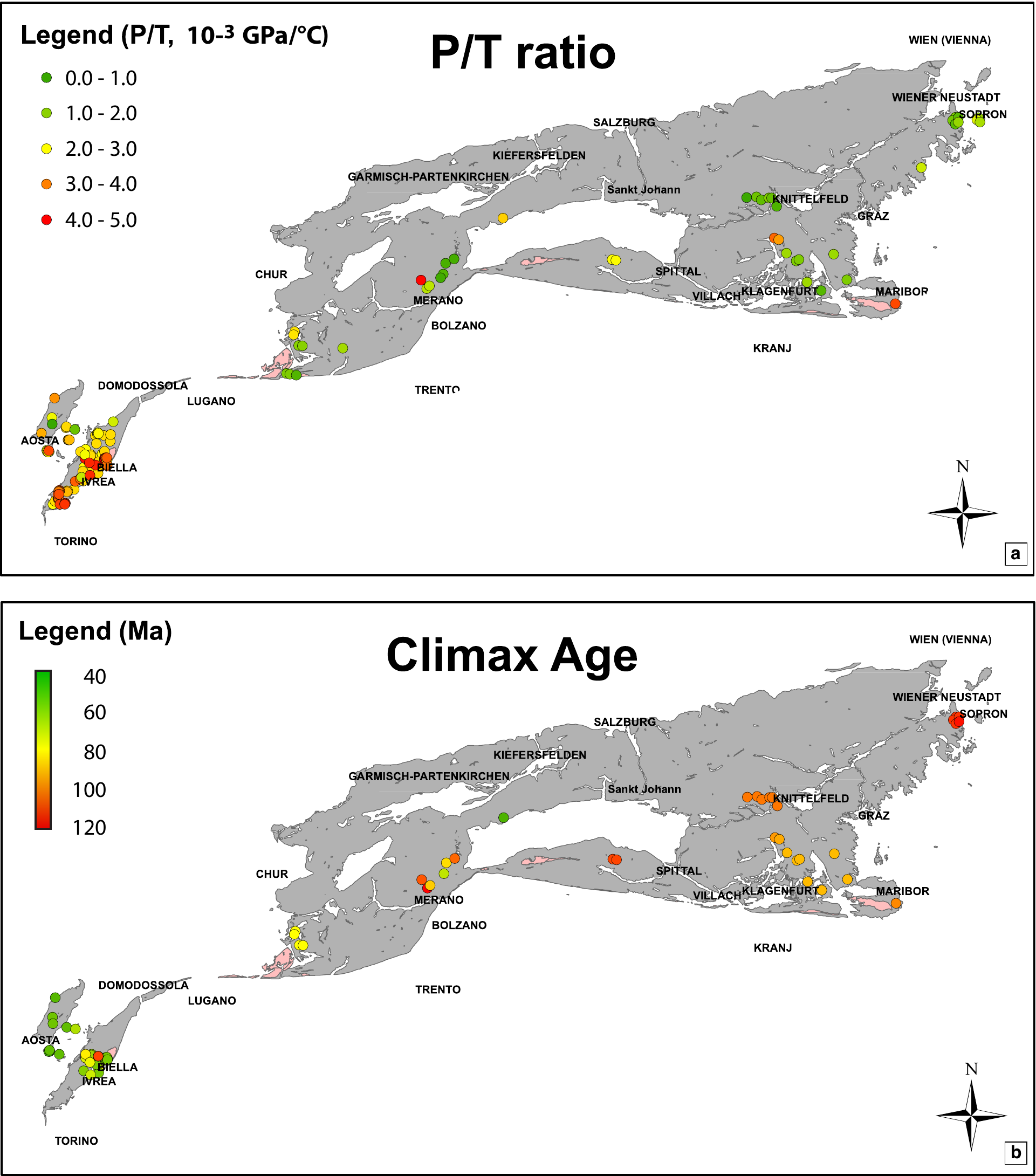}
Fig. 3
\end{figure*}

\vskip 0.5 cm
\noindent
\begin{figure*}[!ht]
\includegraphics[width=14cm]{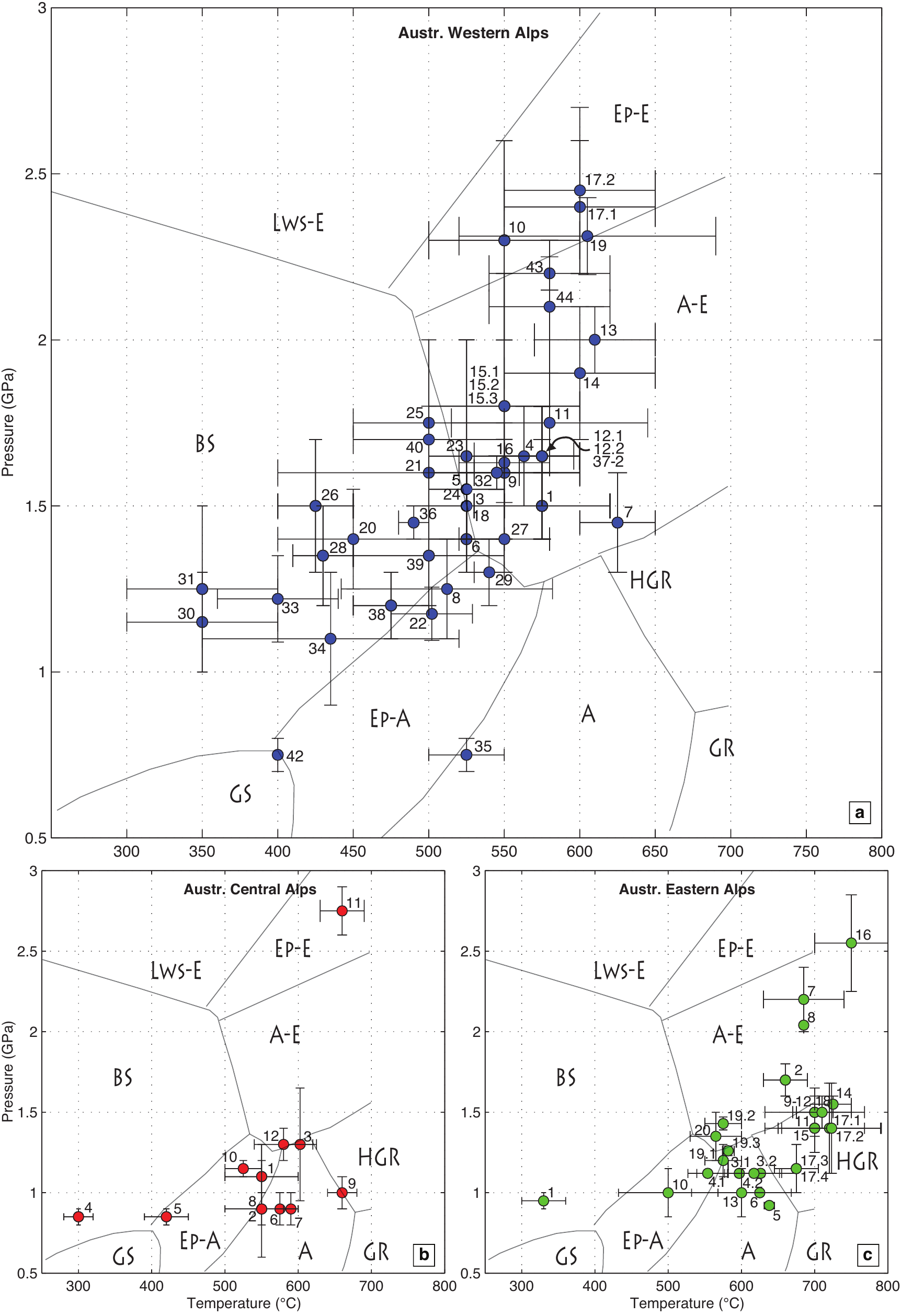}
Fig. 4
\end{figure*}

\vskip 0.5 cm
\noindent
\begin{figure*}[!ht]
\includegraphics[width=14cm]{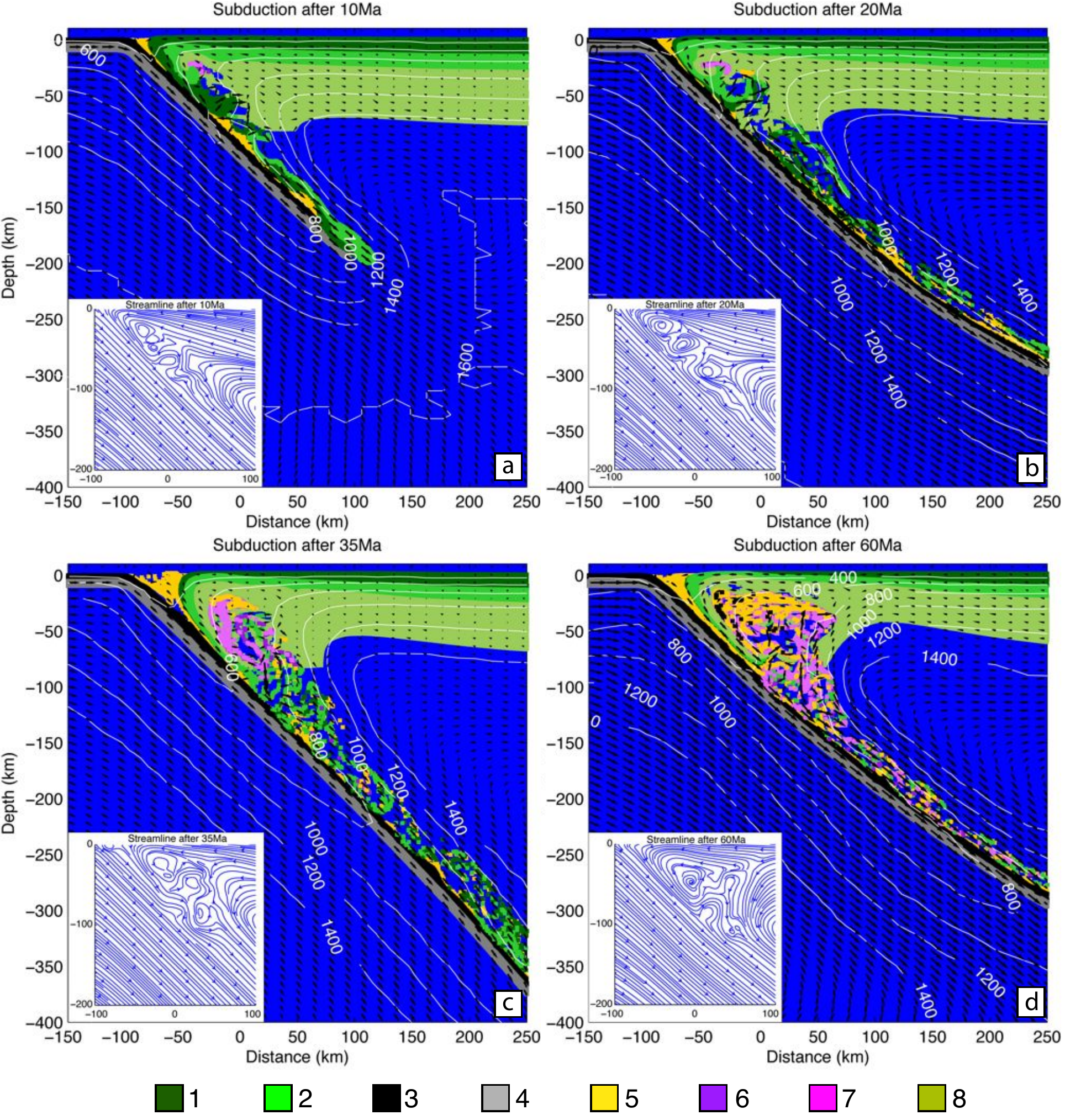}
Fig. 5
\end{figure*}

\vskip 0.5 cm
\noindent
\begin{figure*}[!ht]
\includegraphics[width=14cm]{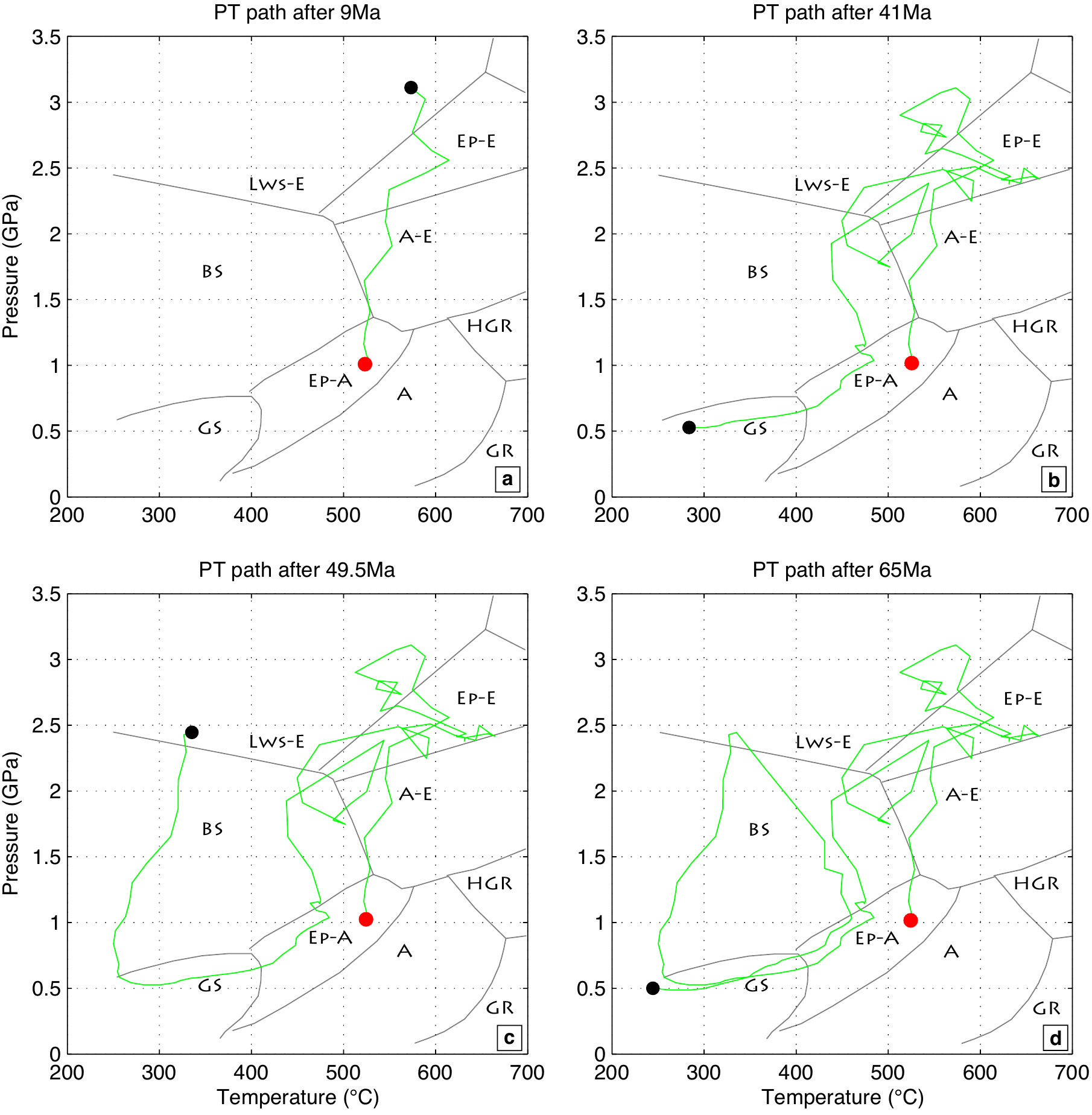}
Fig. 6
\end{figure*}

\vskip 0.5 cm
\noindent
\begin{figure*}[!ht]
\includegraphics[width=14cm]{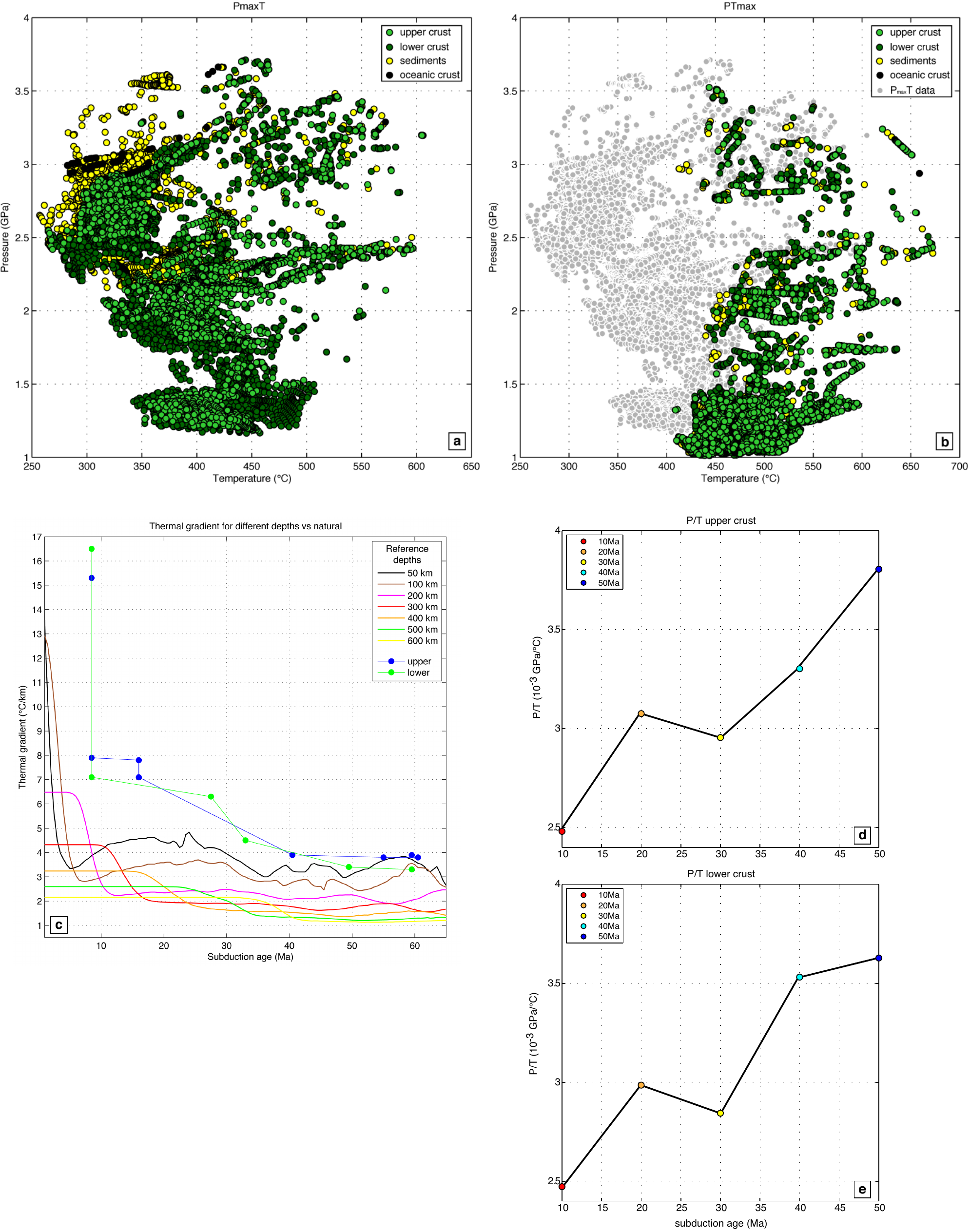}
Fig. 7
\end{figure*}

\vskip 0.5 cm
\noindent
\begin{figure*}[!ht]
\includegraphics[width=14cm]{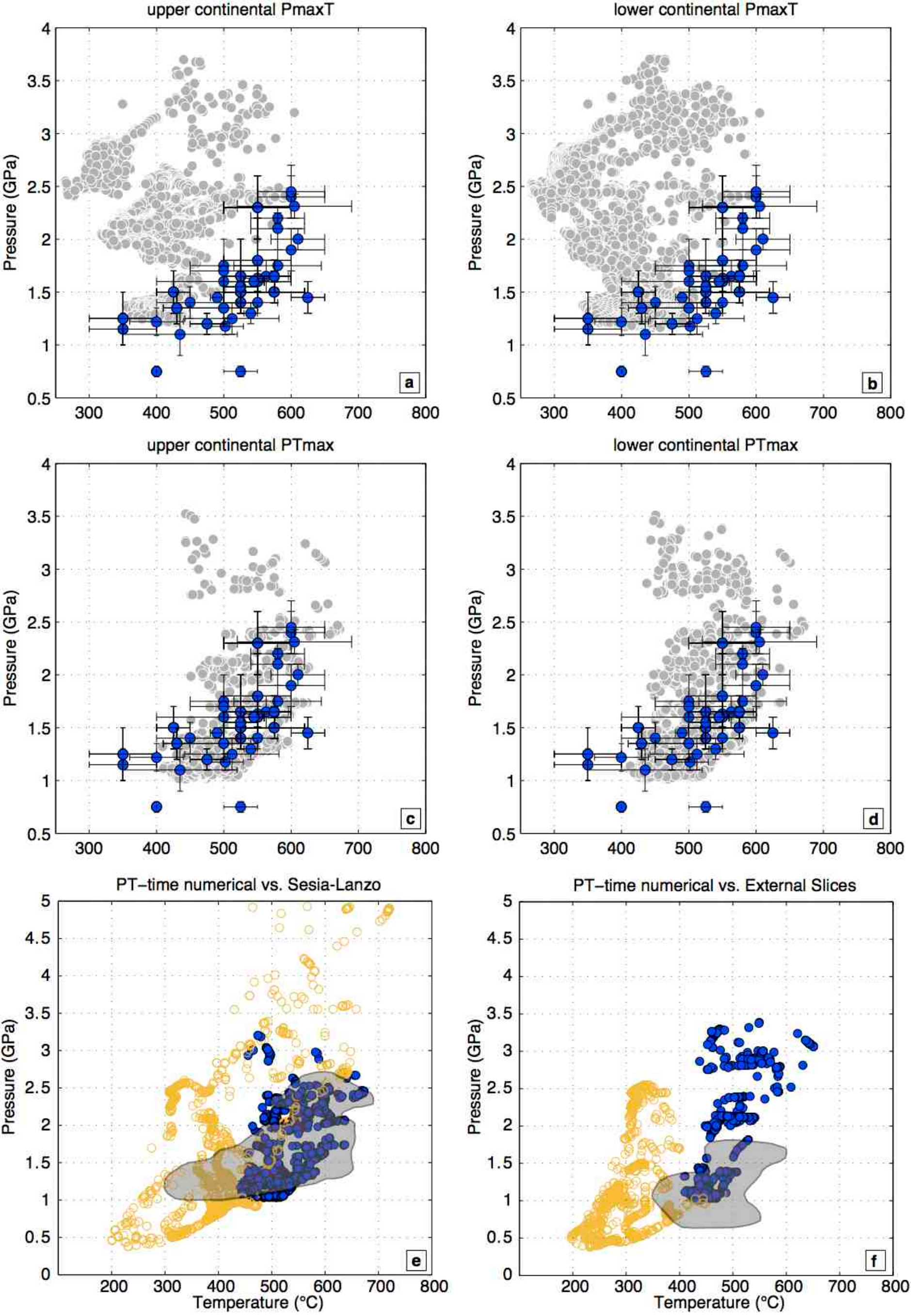}
Fig. 8
\end{figure*}

\vskip 0.5 cm
\noindent
\begin{figure*}[!ht]
\includegraphics[width=14cm]{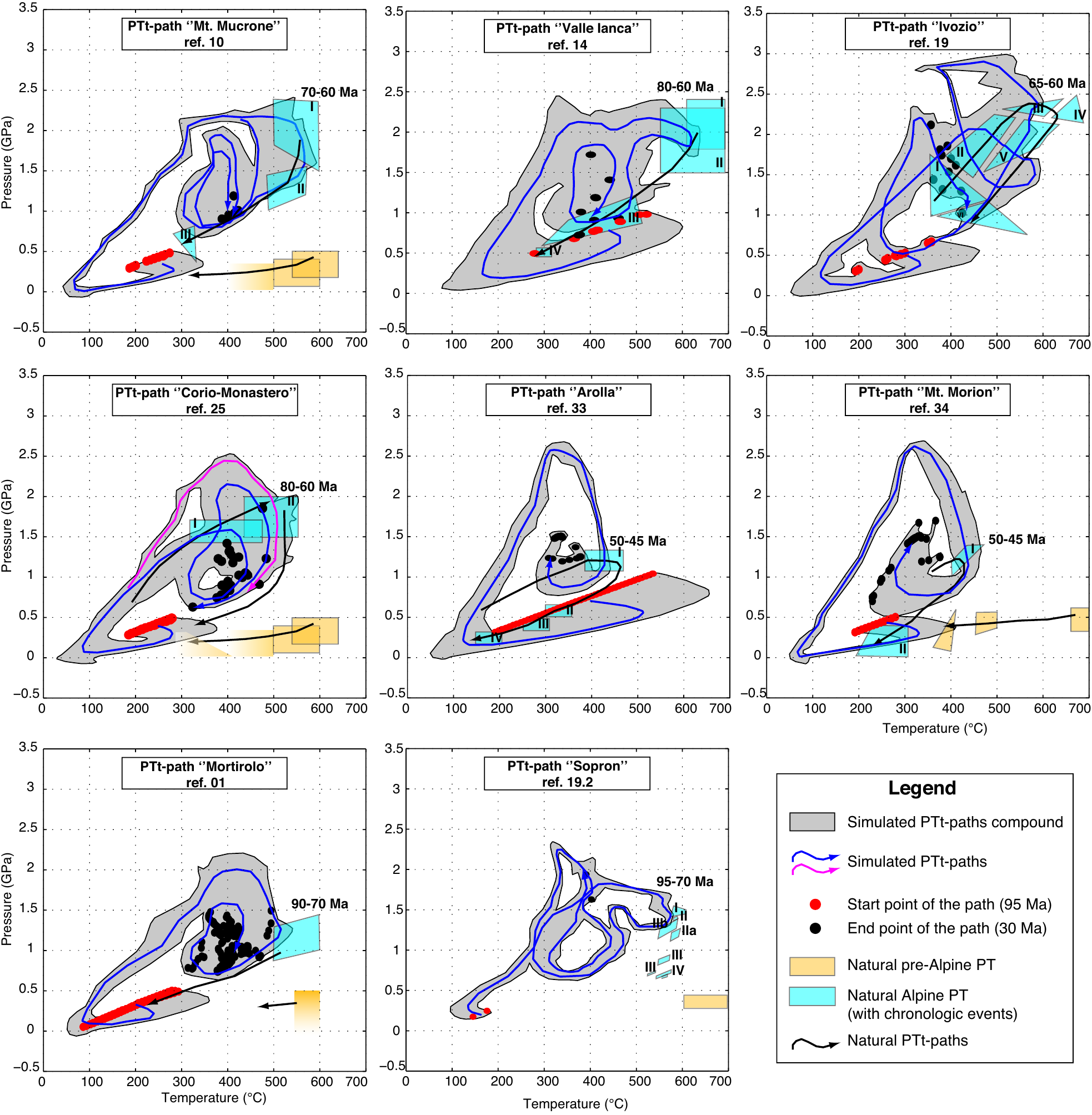}
Fig. 9
\end{figure*}

\vskip 0.5 cm
\noindent
\begin{figure*}[!ht]
\includegraphics[width=14cm]{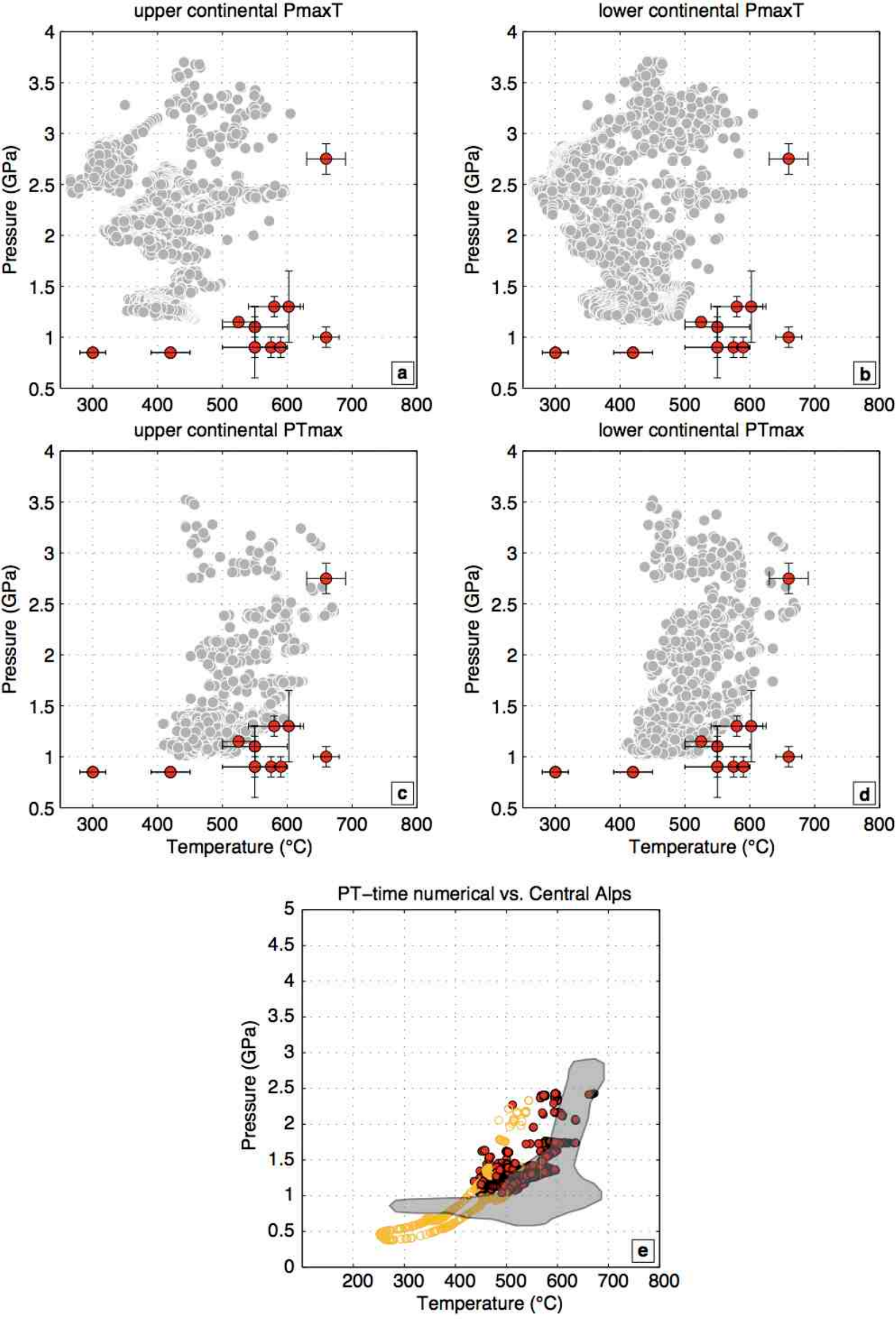}
Fig. 10
\end{figure*}

\vskip 0.5 cm
\noindent
\begin{figure*}[!ht]
\includegraphics[width=14cm]{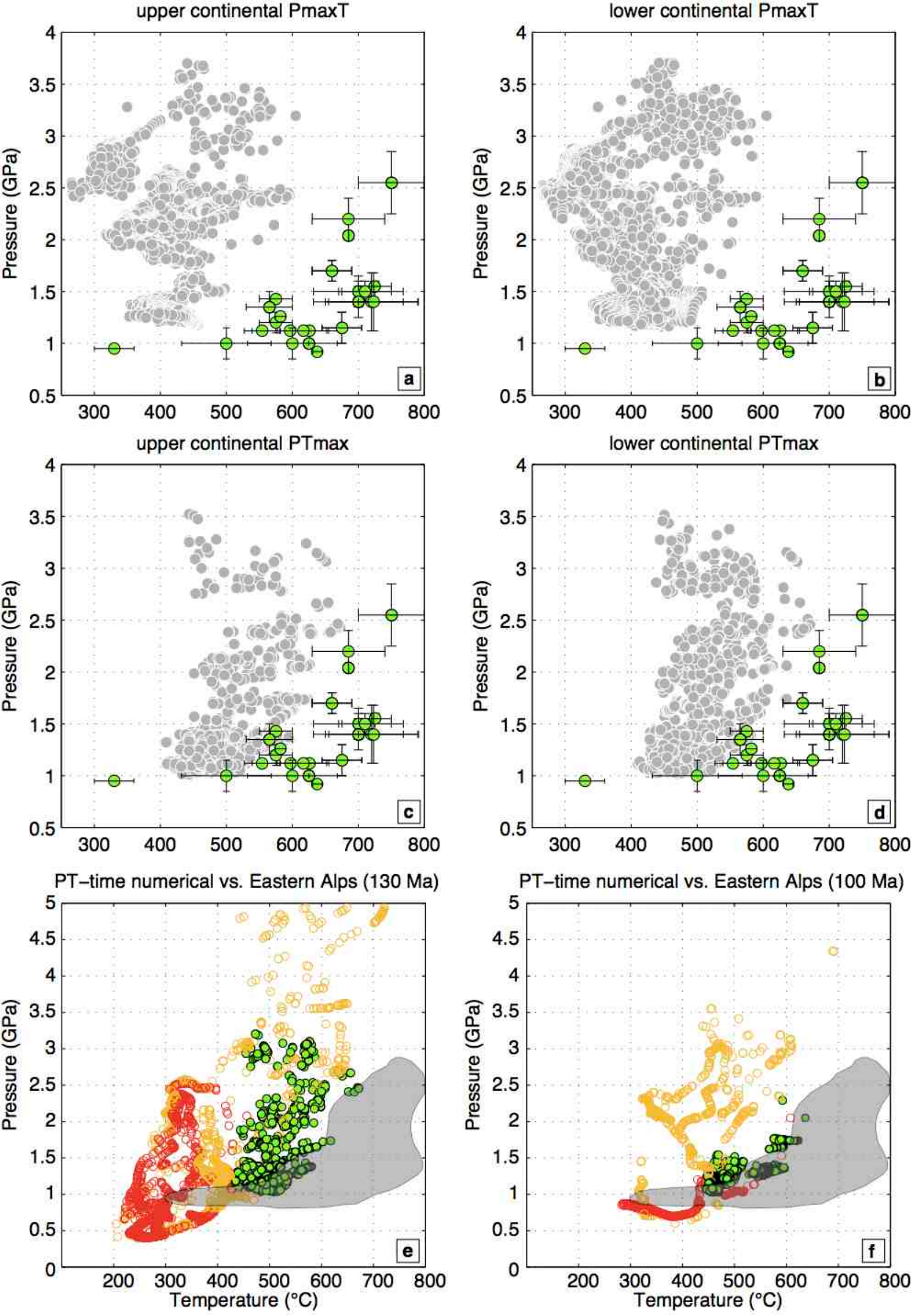}
Fig. 11
\end{figure*}

\clearpage

\begin{table}
\caption{Climax ages for the Austroalpine of the Western Alps: EMC Eclogite Micaschists Complex; GM Gneiss Minuti; II DK II Zona Dioritico-Kinzigitica; RCTS Rocca Canavese Thrust Sheet; CH Chatillon; $DB$ Dent-Blanche; VP Valpelline; RS Roisan; EL Etirol-Levaz; EM Emilius; GL Glacier-Rafray; PK Pillonet Klippe. Keys correspond to the references and locations on Fig. 1a.}
\begin{tabular}{rcccccccccccc}
\hline
\hline
\tiny{Key}&\tiny{Locality}&\tiny{Nappe}&\tiny{Climax}&\tiny{Method}&\tiny{References} \\
	    &                         &\tiny{Met. C.}&\tiny{age (Ma)}& & &\\
\hline
\tiny{4}&\tiny{Lower VdA}&\tiny{EMC}&\tiny{65.0$\pm$5.0}&\tiny{zircon}&\tiny{\citet{Rubatto1999}}\\
\tiny{7}&\tiny{Mt. Mucrone}&\tiny{EMC}&\tiny{71.2$\pm$3.2}&\tiny{Rb/Sr}&\tiny{\citet{Hunziker1974}}\\
            &				&		    &\tiny{118.0$\pm$2.3}&\tiny{Ar/Ar}&\tiny{\citet{Hy1984}}\\
\tiny{8}&\tiny{Mt. Mucrone}&\tiny{EMC}&\tiny{129.0$\pm$15.0}&\tiny{Rb/Sr}&\tiny{\citet{Oberhansli1985}}\\
            &				&		    &\tiny{114.0$\pm$1.0}&\tiny{Rb/Sr}&\tiny{\citet{Oberhansli1985}}\\
\tiny{9}&\tiny{Lower VdA}&\tiny{EMC}&\tiny{69--90}&\tiny{Rb/Sr}&\tiny{\citet{Hunziker1974}}\\
&\tiny{Mt. Mucrone}&&&&\\
\tiny{12.1}&\tiny{Mt. Mucrone}&	\tiny{EMC}&\tiny{65.0$\pm$5.0}&\tiny{zircon}&\tiny{\citet{Rubatto1999}}\\
            &				&		    &\tiny{66--88}&\tiny{allanite}&\tiny{\citet{CenkiTok2011}}\\
\tiny{12.2}	&\tiny{Lower VdA}&\tiny{EMC}&\tiny{65.0$\pm$3.0}&\tiny{zircon}&\tiny{\citet{Rubatto1999}}\\
\tiny{12.3}&\tiny{Cima di Bonze	}&\tiny{EMC}&\tiny{68.0$\pm$7.0}&\tiny{zircon}&\tiny{\citet{Rubatto1999}}\\
\tiny{15.2}&\tiny{Montestrutto}&\tiny{EMC}&\tiny{$\geq$80}&\tiny{Ar/Ar}&\tiny{\citet{Stoeckhert1986}}\\
\tiny{16}&\tiny{Bard}&\tiny{EMC}&\tiny{70--90}&\tiny{Rb/Sr}&\tiny{\citet{Hunziker1974}}\\
\tiny{18}&\tiny{Marine}&\tiny{EMC}&\tiny{$\geq$69.4$\pm$0.7}&\tiny{Ar/Ar}&\tiny{\citet{Ruffet1995}}\\
\tiny{19}&\tiny{Ivozio}&\tiny{EMC}&\tiny{65.0$\pm$3.0}&\tiny{zircon}&\tiny{\citet{Rubatto1999}}\\
\tiny{27}&\tiny{Lower VdA}&\tiny{GM}&\tiny{$\geq$45--60}&\tiny{Rb/Sr}&\tiny{\citet{Hunziker1992}}\\
&&&\tiny{$\geq$60--85}&\tiny{K/Ar}&\tiny{\citet{Hunziker1992}}\\
\tiny{33}&\tiny{Arolla}&\tiny{DB}&\tiny{36.0$\pm$4.0}&\tiny{K/Ar}&\tiny{\citet{Ayrton1982}}\\
	    &                      &               &\tiny{41.0$\pm$0.3}&\tiny{Ar/Ar}&\tiny{\citet{Cosca1994}}\\
\tiny{34}&\tiny{Mt. Morion}&\tiny{DB}&\tiny{46.0$\pm$1.5}&\tiny{Rb/Sr}&\tiny{\citet{Ayrton1982}}\\
\tiny{35}&\tiny{Valpelline}&\tiny{VP}&\tiny{46.0$\pm$1.5}&\tiny{Rb/Sr}&\tiny{\citet{Ayrton1982}}\\
\tiny{37}&\tiny{Etirol-Levaz}&\tiny{EL	}&\tiny{45.0$\pm$0.7}&\tiny{Rb/Sr}&\tiny{\citet{DalPiaz2001}}\\
              &                                &               &\tiny{47.0$\pm$0.9}&\tiny{Rb/Sr}&\tiny{\citet{DalPiaz2001}}\\
\tiny{39}&\tiny{Mt. Emilius}&\tiny{EM}&\tiny{40.0$\pm$0.5}&\tiny{Rb/Sr}&\tiny{\citet{DalPiaz2001}}\\
              &                              &                &\tiny{49.0$\pm$0.5}&\tiny{Rb/Sr}&\tiny{\citet{DalPiaz2001}}\\
\tiny{41}&\tiny{Glacier-Rafray}&\tiny{GL}&\tiny{45.0$\pm$0.4}&\tiny{Rb/Sr}&\tiny{\citet{DalPiaz2001}}\\
\tiny{42}&\tiny{Ayas}&\tiny{PK}&\tiny{74.0$\pm$1.0}&\tiny{Rb/Sr}&\tiny{\citet{Cortiana1998}}\\
\tiny{43}&\tiny{Quincinetto	}&\tiny{EMC}&\tiny{78.5$\pm$0.9}&\tiny{	zircon}&\tiny{\citet{Rubatto2011}}\\
\tiny{44}&\tiny{Brosso}&\tiny{EMC}&\tiny{76.8$\pm$0.9}&\tiny{zircon}&\tiny{\citet{Rubatto2011}}\\
\hline
\hline
\end{tabular}
\end{table}

\clearpage

\begin{table}
\caption{Climax ages for the Austroalpine of the Central-Eastern Alps: UA Upper Austroalpine; LA Lower Austroalpine. Keys correspond to the references and locations on Fig. 1b (Central Alps) and Fig. 1c (Eastern Alps).}
\begin{tabular}{rcccccccccccc}
\hline
\hline
\tiny{Key}&\tiny{Locality}&\tiny{Nappe}&\tiny{Climax}&\tiny{Method}&\tiny{References} \\
	    &                         &\tiny{Met. C.}&\tiny{age (Ma)}& & &\\
\hline
           &                                &                                        &\tiny{Central Alps}&&\\
\hline
\tiny{4}&\tiny{Maloja}	&\tiny{LA/Err}&\tiny{76--89}&\tiny{K/Ar}&\tiny{\citet{Handy1996}}\\
\tiny{5}&\tiny{Engadina}&\tiny{LA/Margna}&\tiny{60--80}&\tiny{Rb/Sr}&\tiny{\citet{Frey1974}}\\
&&&\tiny{60--80}&\tiny{K/Ar}&\tiny{\citet{Frey1974}}\\
\tiny{6}&\tiny{Moos}&\tiny{UA/Schneeberg}&\tiny{84.5$\pm$1.0}&\tiny{Ar/Ar}&\tiny{\citet{Konzett1996}}\\
\tiny{7}&\tiny{Schneeberg}&\tiny{UA/Schneeberg}&\tiny{91.5--96.3}&\tiny{Sm/Nd}&\tiny{\citet{Habler2001a}}\\
\tiny{8}&\tiny{Monteneve}&\tiny{UA}&\tiny{$\geq$76.1$\pm$4.8}&\tiny{Rb/Sr}&\tiny{\citet{Solva2001}}\\
            &                                &\tiny{Texel G.}&\tiny{$\geq$83.2$\pm$2.9}&\tiny{Ar/Ar}&\tiny{\citet{Solva2001}}\\
\tiny{10}&\tiny{Saltaus}&\tiny{UA/Texel G.}&\tiny{143.0$\pm$2.0}&\tiny{Rb/Sr}&\tiny{\citet{Hoinkes1991}}\\
\tiny{11}&\tiny{Ulfas}&\tiny{UA}&\tiny{84.0$\pm$5.0}&\tiny{U/Pb}&\tiny{\citet{Zanchetta2007}}\\
              &                     &\tiny{Texel G.}&\tiny{$\geq$95.0$\pm$5.0}&\tiny{Sm/Nd}&\tiny{\citet{Solva2005}}\\
\tiny{12}&\tiny{Saltaus}&\tiny{UA/Texel G:}&\tiny{85.2$\pm$4.6}&\tiny{Sm/Nd}&\tiny{\citet{Habler2006}}\\
\hline
           &                                &                                        &\tiny{Eastern Alps}&&\\
\hline
\tiny{1}&\tiny{Reckner}&\tiny{LA/Reckner N.}&\tiny{49.5--51.8}&\tiny{Ar/Ar}&\tiny{\citet{Dingeldey1997}}\\
\tiny{2}&\tiny{Schober}&\tiny{UA/Schobergruppe}&\tiny{87--115}&\tiny{Sm/Nd}&\tiny{\citet{Linner1999}}\\
\tiny{3}&\tiny{Mur Valley}&\tiny{UA/Woelz C.}&\tiny{92.4$\pm$14.4}&\tiny{Sm/Nd}&\tiny{\citet{Schuster1999}}\\
\tiny{7}&\tiny{Kupplerbrunn}&\tiny{UA/Saualpe}&\tiny{91.1$\pm$1.3}&\tiny{Sm/Nd}&\tiny{\citet{Thoeni2008}}\\
\tiny{8}&\tiny{Saualpe}&\tiny{UA/Saualpe}&\tiny{90.0$\pm$3.0}&\tiny{Sm/Nd}&\tiny{\citet{Thoeni1996}}\\
\tiny{15}&\tiny{Koralpe}&\tiny{UA/Koralm C.}&\tiny{87.5$\pm$4.5}&\tiny{Sm/Nd}&\tiny{\citet{Miller1997}}\\
\tiny{16}&\tiny{Pohorje}&\tiny{UA/Koralm C.}&\tiny{92.0$\pm$0.5}&\tiny{U/Pb}&\tiny{\citet{Janak2009}}\\
\tiny{17}&\tiny{Sieggraben}&\tiny{UA/Sieggraben C.}&\tiny{$\geq$103.0$\pm$14.0}&\tiny{U/Pb}&\tiny{\citet{Putis2000}}\\
\tiny{18}&\tiny{Sieggraben}&\tiny{UA/Sieggraben C.}&\tiny{$\geq$136--108}&\tiny{Ar/Ar}&\tiny{\citet{Dallmeyer1996}}\\
\hline
\hline
\end{tabular}
\end{table}

\begin{table}
\caption[Material properties used for the simulations]{Material properties used for the simulations. References: (a) \citep{Ranalli1987}, (b) \citep{Kirby1983}, (c) \citep{Chopra1981}, (d) \citep{Karato1993}, (e)
\citep{Dubois1997,Best2001}, (f) \citep{Rybach1988}, (g) \citep{Gerya2002}.}
\begin{tabular}{llccccccc}
\hline
\hline
{\tiny Materials} & {\tiny Rheology} &{\tiny $\mu^0$($Pa s^{-1}$)}  &{\tiny n}&{\tiny $\rho_0$($kg/m^3$)} & {\tiny k(W/mK) }  & {\tiny   $H_r$($\mu$W/$m^3$)}    &{\tiny E(KJ/mol)} &{\tiny Refs.}\\
\hline
{\tiny Continental crust} &{\tiny dry granite} &{\tiny $3.47\cdot10^{21}$}&{\tiny3.20 }&{\tiny 2640} &{\tiny 3.01} &{\tiny 2.50} &{\tiny 123}&{\tiny a,e,f}\\
{\tiny Oceanic crust: upper}&{\tiny $10^{19}$ }&  &&{\tiny 2961} &{\tiny 2.10} &{\tiny 0.40} &{\tiny 260}&{\tiny b,e,f}\\
{\tiny Oceanic crust: lower}&{\tiny diabase }&{\tiny$1.61\cdot10^{22}$} &{\tiny3.40 }&{\tiny 2961} &{\tiny 2.10} &{\tiny 0.40} &{\tiny 260}&{\tiny b,e,f}\\
{\tiny Sediments} &{\tiny $10^{19}$} && &{\tiny 2640} & {\tiny 3.01 }&{\tiny 2.50} &{\tiny 123}&{\tiny e,f}\\
{\tiny Dry mantle} &{\tiny dry dunite} &{\tiny $5.01\cdot10^{20}$}&{\tiny3.41} &{\tiny 3300}&{\tiny 4.15} &{\tiny 0.002} &{\tiny 444} &{\tiny c,e,f}\\
{\tiny Serpentinized mantle} &{\tiny $10^{19}$ }& & &{\tiny 3000} & {\tiny4.15} &{\tiny 0.002 }& {\tiny444}&{\tiny e,f,g}\\
{\tiny Water} &{\tiny $10^{23}$ }      & & &{\tiny 1000 } &{\tiny 0.60 }&            &\\
{\tiny Atmosphere} &{\tiny $10^{23}$ }      & & &{\tiny 1.18 } &{\tiny 0.03 }&            &\\
\hline
\hline
\label{tab:mat_parameters}
\end{tabular}
\end{table}

\end{document}